\documentclass[12pt,a4]{article}
\usepackage[T1]{fontenc}
\usepackage{lmodern}
\usepackage[utf8]{inputenc}
\usepackage{setspace}
\usepackage{etoolbox}
\makeatletter
\patchcmd{\appendix}{\@Alph}{\@Roman}{}{}
\makeatother

\usepackage{amsmath}
\usepackage{amssymb}
\usepackage{amsthm}
\usepackage{mathtools}
\usepackage{mathrsfs}
\usepackage{breqn}

\usepackage[margin=1in]{geometry}
\usepackage{enumerate}
\usepackage{enumitem}
\setlist[enumerate,1]{label=(\arabic*)}
\setlist[itemize,1]{label=--}    
\usepackage[dvipsnames]{xcolor}
\usepackage{hyperref}
\usepackage{float}
\usepackage{indentfirst}

\usepackage{sectsty}
\sectionfont{\centering}
\subsectionfont{\centering}

\usepackage{tikz}
\usetikzlibrary{decorations.pathreplacing}
\usepackage{mathtools}
\usetikzlibrary{positioning}
\usetikzlibrary{trees}
\usepackage{graphicx}

\newcommand{\bb}{\mathbb}

\newcommand{\U}{\bigcup}

\newcommand{\und}{\underline}
\newcommand{\lims}{\lim\limits}

\newcommand{\mcal}{\mathcal}

\newcommand{\var}{\text{var}}

\newcommand{\supp}{\text{supp}}
\renewcommand{\epsilon}{\varepsilon}

\newcommand{\blue}[1]{\color{blue}#1 \color{black}}

\newtheorem{theorem}{Theorem}
\newtheorem{lemma}{Lemma}
\newtheorem{definition}{Definition}
\newtheorem{proposition}{Proposition}

\newtheorem{assumption}{Assumption}

\newtheoremstyle{remboldstyle}
  {}{}{}{}{\bfseries}{.}{.5em}{{\thmname{#1 }}{\thmnumber{#2}}{\thmnote{ (#3)}}}
\theoremstyle{remboldstyle}

\DeclareMathOperator*{\argmax}{\arg\!\max}
\DeclareMathOperator*{\argmin}{\arg\!\min}
\DeclareTextFontCommand{\emph}{\slshape}

\usepackage{natbib}
\nocite{*}

\title{Reputation in the Shadow of Exit}

\author{Daniel Luo\footnote{Department of Economics, Massachusetts Institute of Technology. daniel57@mit.edu.
\\  This paper strictly supersedes ``Reputation and Risk in Regimes,'' and ``Reputation in Repeated Global Games of Regime Change,'' which originated from work done when I was an undergraduate at Northwestern University. I am indebted to Drew Fudenberg, Stephen Morris, and Alex Wolitzky for guidance. I thank George-Marios Angeletos, Nina Fl\"uegel, Eric Gao, Andrew Koh, Andrew Komo, Anna Merotto, Alessandro Pavan, Harry Pei, Larry Samuelson, Alvaro Sandroni, Udayan Vaidya, Muhamet Yildiz, and Alison Zhao for helpful feedback regarding this project. I also thank participants of MIT Theory Lunch, Stonybrook 2024 and the 2023 Carroll Round for comments. Coarse.ink and Refine.ink were used to check this paper for clarity. This research was partially supported by the NSF Graduate Research Fellowship and the Northwestern University Baker program in undergraduate research. All errors are mine alone.}}
\date{\today}

\begin{document}

\maketitle

\begin{abstract}
    I study reputation formation when player actions endogenously determine the probability the game permanently ends. 
    Exit can render reputation useless even when (1) players do not discount the future and (2) the Stackelberg payoff is large, in contrast to canonical commitment payoff theorems. 
    However, I identify conditions for the long-run player to attain at least their Stackelberg payoff in all Markov equilibria. Along the way, I highlight the role Markov strategies play in pinning down the value of reputation. I apply my results to give qualified commitment foundations for the infinite chain-store game. 
    I also analyze repeated global games with exit, and obtain new predictions about regime survival. 
\end{abstract}



\newpage
\onehalfspacing
\section{Introduction}
Reputation has long been recognized as a powerful way to refine predictions in repeated games. By endowing the long-run player with the ability to imitate a commitment type who always takes some action $a$, a rational long-run player can build a reputation for playing $a$, and thus secure the payoff they would obtain \textit{as-if} they were known to be committed to that action. 
Successful reputation formation relies on two important features of the repeated game: First, that the game itself is repeated sufficiently often, so that the long-run player can undertake the commitment action enough times to build their reputation, and second, that the long-run player is patient relative to the speed of reputation formation, so that the ``reputation-building'' phase of the game contributes little to their average payoff. 

Yet there are a variety of settings where these assumptions may fail to hold. 
Consider, for example, the chain store game of \cite{KrepsWilson82} and \cite{MilgromRoberts82} (hereafter, simply ``chain store game'' or ``KWMR''). A crucial assumption in their setup is that the incumbent is able to repeatedly maintain their position in an a-priori fixed number of markets, even if they unsuccessfully deter entry. Yet in many incumbent-entrant games, including the chain store game, successful entry by a challenger can displace the incumbent, changing the structure of the game in future periods, and either forcing the incumbent to exit or the game to prematurely end.\footnote{For example, the monopolist loses their market power in the KWMR chain store game if a challenger successfully enters and is accommodated.} Similarly, in the product choice game of \cite{mailath2006repeated} (Chapter 15.1) and \cite{pei2020interdependent}, it may be that a long enough string of low quality forces the firm to exit. 
Finally, in models of repeated regime change (c.f. \cite{Huang17}), a successful currency attack could end the current regime, i.e. lead to the toppling of the central bank. Players, realizing this, may change their behavior, which could in turn affect reputational dynamics. 

In such games, the assumption that the stage game will be infinitely repeated may be unreasonable, as player actions can lead to premature permanent exit. Consequently, the expected number of periods play persists may be much shorter than the ``upper bound on the number of times the short-run players can fail to play as Stackelberg followers,'' (c.f. \cite{FudenbergLevine89}) blocking both the long-run player's ability to build a reputation and for the early stages of the game to be unimportant for their average payoff. 
Indeed, absent potentially restrictive assumptions on the exit technology itself, little can be said about reputation formation in the shadow of exit, restoring the indeterminacy of the complete-information folk theorem (see Proposition \ref{p: exit hurts reputation}). Despite the potential appropriateness of directly modeling exit in several of the leading examples studied by the reputation literature, little has been said about the effect exit has on incentives for reputation formation. 

This paper studies reputation formation in repeated games where player actions endogenously determine if the game permanently ends. Formally, I consider a model where many short-run players (choosing $a$) interact with a long-run player (choosing $c$) who can be either a rational type or a single committed type who plays a Stackelberg action. In addition to payoffs, the action profile $(a, c)$ also jointly determines a \emph{survival probability} the game continues; with complementary probability, the game permanently ends. 
The rational type wishes to maximize their expected payoff by potentially imitating the committed type. Crucially, fear of exit can shut down traditional incentives for reputation formation, as the effective discount rate given short-run player actions can be bounded away from $1$ even absent long-run player time discounting. Thus there is no guarantee that investing in the Stackelberg action today will be rewarded with short-run Stackelberg best replies with high frequency in the future, rendering standard arguments from the reputation literature inapplicable. 

The main result of this paper identifies two jointly sufficient (and necessary, in \emph{monotone} games) conditions  that together are enough for reputation effects to select the long-run player's Stackelberg payoff in all Markov equilibrium: 

\begin{enumerate}
    \item \emph{Incentive compatibility}: the ``effective'' discount factor is high enough that playing the Stackelberg action can be enforced in some equilibrium of the repeated game. 
    \item \emph{Markov beliefs}: whenever short-run players hold identical beliefs about the long-run player's rationality, they act the same way. 
\end{enumerate}

The incentive compatibility constraint requires that, at any feasible short-run player best response, the action $c^*$ is enforceable, and is taken as hypothesis throughout the analysis of the paper. 
The Markov beliefs condition is captured by a Markovian refinement that short-run player actions are measurable with respect to their belief about the long-run players' type. This ensures that short-run players act on their beliefs about the long-run player's rationality, and rules out equilibria where the long-run player exits with high probability in the first period due to a lack of variation in continuation play in the future. Indeed, I prove in the absence of this condition an \emph{anti-commitment payoff theorem} (Proposition \ref{p: exit hurts reputation}) by constructing equilibria which guarantee the long-run player an arbitrarily low payoff absent the Markov refinement, even when payoffs satisfy my incentive compatibility condition. 

Theorem \ref{t: commitment theorem} leads to novel comparisons of Markov equilibria. In the complete information game, Markov equilibrium cannot sustain the Stackelberg payoff even when it is an equilibrium of the repeated game in general (Propositions \ref{p: folk theorem} and \ref{p: markov folk theorem}). However, when the long-run player faces reputational incentives, the long-run player plays their Stackelberg action at every period and perfectly pools with the committed type in all Markov equilibria.\footnote{This contrasts with past work, which shows that even under perfect monitoring there can be a continuum of equilibria, see \cite{PeiLi21}.}
Thus, the Markov refinement not only helps select the long run player's best equilibrium payoff in the perturbed reputation game, but selects a payoff for the long-run player which is strictly higher than their best Markov payoff in the complete information game without reputational incentives. 

I illustrate my results by returning to the chain store game. I show that when accommodating the entrant leads to some probability the game permanently ends, one of two things can happen depending on whether the Markov beliefs restriction is adopted. First, in Markov equilibrium, there exists a unique Markov equilibrium profile which secures the long-run player their Stackelberg payoff. However, in sequential equilibrium, the long-run player's payoff may secure only their minimax payoff, exactly because of the existence of equilibria where short-run players vary play in ways that are payoff-irrelevant. That the long-run player cannot secure a high payoff in sequential equilibrium contrasts sharply with the results of the chain store game absent exit. I interpret these results as giving qualified commitment foundations for the chain store game with exit. 

Finally, I apply my model to repeated globalized regime change games with exit. A long-run player endogenously determines, at a cost, some coordination friction, after which short-run players see noisy signals about the regime's strength and decide to attack. The game continues if and only if sufficiently few players attack that the regime is able to overcome the revolution in that period. Such a setup is similar to the game in \cite{AngeletosHellwigPavan07}, but with reputational effects and endogenous regime actions. I use Theorem \ref{t: commitment theorem} to establish an asymptotic selection result,\footnote{I consider the limit where both the idiosyncratic and aggregate noise vanish in order to satisfy my incentive compatibility condition.} Theorem \ref{t: global games theorem}: In the unique Markov equilibrium, the regime attains their Stackelberg payoff if and only if the average strength of the regime is above the \emph{endogenously chosen} risk-dominant level. 
Conversely, if the average strength is below the risk-dominant level, then the regime dies almost surely in the first period and attains an arbitrarily low payoff. Since observed equilibrium play jumps at the risk-dominant threshold, my result sheds light on why seemingly ex-ante similar regimes can differ dramatically ex-post in both (1) observed punishment and (2) length of survival. 

The remainder of this paper is organized as follows. Section 1.1 reviews the related literature. Section 2 introduces the model. Section 3 establishes the main results of the paper. Section 4 applies the model to the chain-store game. Section 5 extends the framework to study repeated global games of regime change. Section 6 concludes. Omitted proofs are collected in the appendix. 
A supplemental appendix discusses the discounted repeated game.  

\subsection{Related Literature}

This paper contributes to the extensive literature on reputation formation in repeated games, started by \cite{KrepsWilson82} and \cite{MilgromRoberts82}. I adopt the canonical framework of \cite{FudenbergLevine89} and \cite{FudenbergLevine92}, who establish the generality of reputational effects in a wide class of infinitely repeated games with long-run and short-run players (see \cite{FudenbergKrepsMaskin90} and \cite{fudenberg1994efficiency}). There is a large and rich literature studying variations to their model; see Part IV of \cite{mailath2006repeated} or \cite{mertens2015repeated} for a survey.

This paper is also related to \cite{pei2024reputation} and \cite{community2025enforcement}, who study endogenous records as a key friction to reputation building. Importantly, \citeauthor{community2025enforcement} shows that endogenous records accord differing roles to the exit probability and patience of players, and consequently there is rich asymptotic behavior depending on whether discounting or exit vanishes first. In contrast, I endogenize the exit probability instead of records, so in particular allow it to always be bounded away from its patient limit. This leads to a different exercise than \citeauthor{pei2024reputation}'s models, and thus different predictions despite a similar decomposition of the exit probability.  
This paper also relates to \cite{PeiLi21}, who characterize equilibrium play in the \cite{FudenbergLevine89} model without exit; I characterize unique Markovian play with exit. 

I also relate to a smaller literature on strategy-contingent discounting in repeated games without reputation effects. \cite{Koppel2004Strategy} studies the prisoner's dilemma in a complete information game with endogenous discounting, while \cite{NeilsonWinter1996Endogenous} give a general folk theorem when the discount rate is endogenously chosen by one player. When player payoffs are not time-separable, including as a special case the setting where the discount rate is endogenous, \cite{KochovSong2023Intertemporal} and \cite{KochovSong2025Unobservable} present a general folk theorem. I contribute to this literature by (1) studying a specific form of time endogenous discounting---permanent exit in the game, and (2) analyzing the perturbed game with reputation effects, which until now this literature has not studied. 

Relatedly, \cite{abreu2000bargaining} and the literature on reputational bargaining (see \cite{Fanning2022} for a review) assume the game ends when one player concedes, a particular form of exit. While the way endogenous exit is encoded is related, the models are quite different: for example, bargaining models are zero-sum, assume all players are long-run, and have very special exit structures.
Further afield, \cite{BarIsaac2003} study reputation effects in a monopoly selling model with interdependent values and characterize when the (good) monopolist always sells; their approach to reputation is distinct from the classical \citeauthor{FudenbergLevine89} approach and they focus on equilibrium behavior instead of payoffs. 
That these models are economically different can be seen in the new technical challenges that arise in my model and the distinctiveness of the qualitative conclusions that attain. 

My paper also relates to models of ``bad reputation'' in extensive form reputation games (see \cite{ElyValimaki03}, \cite{ely2008when}, and \cite{luo2024marginal})  where a shutdown of statistical learning is similar to exit. I contribute to this literature by first directly working with the exit probability as a primitive instead of a derived object due to monitoring, and show that under this interpretation there are conditions where reputation can once again secure a high payoff for the long-run player. 

Using Markov refinements to make predictions in repeated games with reputation effects is not new, and I see my refinement as contributing to a long and historical strand of literature that makes this assumption in order to say something meaningful about equilibrium play. For example, \cite{Pei2016} and \cite{LukyanovShamrukLogina2025} use a Markov refinement to establish ``inverse U-shaped'' behavior in labor and goods markets, respectively, to reputational incentives, while \cite{KostadinovRoldan2025} focus on Markov equilibrium to characterize the dynamics of optimal monetary policy with reputational concerns. \cite{GiannitsarouToxvaerd} use Markov refinements to render the repeated global game (which is a leading application of my framework) tractable and make predictions. 
\cite{BhaskarMailaithMorris} provide a foundation for Markov perfection in long-run short-run games with bounded memory: they are the only purifiable equilibrium. In games with reputation, \cite{SannikovFaingold} show that continuous-time games select a unique sequential equilibrium which is Markovian in the belief of the long-run player's rationality---exactly my criteria. My paper gives an alternative justification for the Markov refinement, as a minimal way of selecting payoffs in the repeated game with endogenous exit. 

Finally, the regime change application builds on the rich literature of equilibrium selection in global games, started by \cite{CvD93} and \cite{morris1998unique} (see \cite{MorrisShin03} for a survey). The model's key innovation in this area is to (1) allow the regime to choose, at a cost, the optimal punishment level, and (2) embed the game into a repeated regime change game reminiscent of \cite{AngeletosHellwigPavan07} and \cite{Chassang10} but with an i.i.d. state.  The first change relates to a large literature in global games where some ``large'' player can take an action that modifies the global game and leads to new applied insights about regime change (see \cite{AngeletosHellwigPavan06}, \cite{Edmond13}, and \cite{morris2024repression}). Proposition \ref{p: patience effect} makes an independent contribution in this direction by establishing a novel comparative static that rationalizes why stronger or more patient regimes are more likely to be cruel, even when doing so harms the regime. The second change allows me to analyze reputational effects, and unlike \cite{AngeletosHellwigPavan07} and \cite{AngeletosPavan13}, establish the existence of a unique Markov equilibrium.

\section{Model}
\subsection{The Stage Game}
There are $N + 1$ players, $i \in \{L\} \cup [N]$, representing a long-run player $L$ who interacts in each period with $N$ short-run players. Each short-run player $i$ chooses an action from a compact metrizable set, $a_i \in A_i$; let $a \in \prod_i A_i = A$ denote a profile of short-run player (pure) actions. The long-run player chooses an action $c \in C \subset \bb{R}$, where $C$ is a countable (potentially finite) set including $0$.\footnote{In the global games application, the set of actions will be a compact interval. We justify this by noting that if $g: Q \to \bb{R}$ is a restriction of a continuous function $g: X \to \bb{R}$, where $Q \subset X$ is countable and dense, there is a unique continuous function $\overline g: X \to \bb{R}$ such that $\overline g|_{Q} = g$. However, the countability assumption helps significantly when defining equilibrium beliefs off-path, so I maintain it throughout the general analysis.}
The entire action profile also generates a finite signal $Y$, drawn according to a distribution $\rho(\cdot | a, c)$.\footnote{Short-run player monitoring is irrelevant to the analysis and included only for generality.}
The long-run player's action is perfectly monitored.

The action profile $(a, c)$ jointly determines a \emph{survival probability}, represented by a function $f: A \times C \to (0, 1)$. This can be interpreted as the probability that the game continues. The complementary probability $1 - f(a, c)$ is the \emph{exit probability.} 

The long-run player's expected stage game payoff is thus given by $f(a, c)v(a, c)$, where $v(a, c)$ is the payoff they attain conditional on surviving. Note here we assume that the long-run player only obtains their payoff if they survive; this makes the analysis more convenient (particularly Proposition \ref{p: exit hurts reputation}), and simplifies the representation of the payoff streams, but does not substantially change any of the results. Short-run player $i$ has payoffs $u_i(a, c)$.\footnote{Modifying short-run player payoffs to be $f(a, c)u_i(a, c)$ will not affect any of the analysis, though it will implicitly change which actions are best replies and thus implicitly affect enforceability constraints.}
Suppose $(u, v, f)$ are all bounded and continuous in all arguments. Throughout, assume payoffs are weakly positive. Moreover, assume that $f(a, c) \leq 1 - \xi$ for some $\xi > 0$ always. This can be interpreted as saying there is always some (arbitrarily small) probability exit occurs regardless of what actions are taken.\footnote{This ensures the undiscounted payoff streams are well-defined, and in particular converge.}

For any (potentially mixed) long-run player action $\gamma \in \Delta(C)$, denote by $\mcal N(\gamma)$ the Nash correspondence at $\gamma$ over all strategies; that is, 
\[ \mcal N(\gamma) = \left\{\alpha \in \prod_i \Delta(A_i) : \supp(\alpha_i) \subset \argmax_{\overline a_i \in A_i} \bb{E}_{a_{-i} \sim \alpha_{-i}}\bb{E}_{c \sim \gamma}[ u_i(\overline a_i, a_{-i}, c) ]  \right\}. \]
Note that $\mcal N$ has closed graph and is compact-valued, so in particular is upper hemi-continuous. Throughout, I assume that $\mcal N(\beta c^*)$ is connected, where $\beta c^*$ places probability $\beta$ on $c^*$ and $1 - \beta$ on $0$\footnote{This is automatically satisfied whenever $\mcal N$ is single-valued and continuous, or with a single short-run player, as in both applications.}.
When there is no ambiguity, I abuse notation to (1) use $f(\alpha, \gamma)v(\alpha, \gamma)$ to mean the expected payoff given lotteries $\alpha \in \Delta(A)$, $\gamma \in \Delta(C)$,  and (2) identify actions $a, c$ with degenerate lotteries in $\Delta(A)$, $\Delta(C)$ respectively. 

Throughout the analysis, it will be useful (expositionally) to impose the assumption the long-run player has a dominant stage game action, labeled at $c = 0$.

\begin{assumption}
\label{ass: dominance}
    For all $a$ and $m \in \bb{R}_+$, $\{0\} = \argmax_{c \in C} f(a, c)[v(a, c) + m] = \argmax_{c \in C} v(a, c)$. 
\end{assumption}

The first part of the assumption is a dominance requirement that simplifies the ``worst punishment'' from deviating from a prescribed equilibrium. In particular, when $m = 0$, it implies the set of one-shot Nash payoffs attainable in the one-shot game can be represented as $\{f(\alpha, 0)v(\alpha, 0)\}_{\alpha \in \mcal N(0)}$. 
Towards this end, let $\overline V_0 = \max_{\alpha \in \mcal N(0)} \frac{f(\alpha, 0) v(\alpha, 0)}{1 - f(\alpha, 0)}$ denote the long-run player's largest payoff from playing the static optimum repeatedly, and analogously let $\und V_0$ be their smallest such payoff. When $m > 0$, Assumption \ref{ass: dominance} says that $0$ remains optimal even when there is continuation value in the future, and helps discipline the set of Markov equilibria in the complete information game.
When $f$ is independent of $c$ and strictly positive, (i.e. depends only on $a$, as is the case for both of our applications), Assumption \ref{ass: dominance} is satisfied so long as $\{0\} = \argmax_{c \in C} v(a, c)$. 

\subsection{The Repeated Game}
The long-run player plays the stage game in each of infinitely many periods ($t = 0, 1, 2, \dots$), weighting each period equally (conditional on not exiting). In each period, they face a population of $N$ short-run players who take myopic best replies. 

Time-$t$ histories are elements $h^t = (y_s, c_s)_{s = 0}^{t - 1}$ of past short-run player signals and long-run player actions. Let $H^t$ denote the set of time-$t$ histories, $H = \U_t H_t$ the set of all histories, and $H^\infty = (Y \times C)^\infty$ the set of all infinite histories.\footnote{Throughout, I will sometimes drop the dependence on $Y$ when referring to a history simply as an element of $C^\infty$; this is meant to be read as an equivalence class of all histories with that realization of long-run player actions, for any accompanying realization of public signals $Y$.}
Short-run player strategies are functions $\sigma_i: H \to \Delta(A_i)$, while the (rational) long-run player's strategy is a function $\sigma_L: H \to \Delta(C)$.\footnote{Following \cite{mertens2015repeated}, there is a canonical mapping from strategies $\sigma: H \to \Delta(\mcal X)$ and strategies $\sigma: H^\infty \to \Delta(\mcal X)$ for $\mcal X \in \{C, A\}$. When it is useful (i.e. when topologizing $\sigma$), I will use these two codomains interchangeably.} Throughout, I assume players have access to a public randomization device. 

Given a sequence of actions $\{a_t, c_t\}_{t = 0}^\infty$, the long-run player's payoff is 
\[ \mcal V(\{a_t, c_t\}) = \sum_{t = 0}^\infty \left( v(a_t, c_t) \prod_{s = 0}^t f(a_s, c_s)\right). \] 
Note payoffs here are not discounted. \blue{\ref{Appendix B}} studies the discounted repeated game and shows the game here is the standard limit as players become patient, but with exit. However, discounting payoffs does not add anything to the analysis and would introduce unnecessary notation. Consequently, I omit the discount rate in the main exposition, and refer to this as the case of ``extreme patience'' for the long-run player. 

Define the long-run player's lower Stackelberg payoff as  
\[ V_* = \sup_{c \in C} \min_{\alpha \in \mcal N(c)} \sum_{t = 0}^\infty f(\alpha, c)^{t + 1}  v(\alpha, c) = \sup_{c \in C} \min_{\alpha \in \mcal N(c)} \frac{f(\alpha, c)v(\alpha, c)}{1 - f(\alpha, c)} \]
and let $c^*$ be a maximizer in $V_*$.\footnote{Throughout, suppose such a maximizer exists. This is true if, for example $C$ is finite.} Let $V^*$ be the upper Stackelberg payoff, defined analogously. 
Payoffs are written such that exit is permanent: once exit has occurred, the long-run player leaves the game, and hence their payoff is zero in all future periods regardless of any hypothetical actions that would have been taken. 

The long-run player can either be a \emph{rational type}, $\omega_R \in \Omega$, who seeks to maximize their discounted payoff, or a \emph{committed type}, $\omega_c \in \Omega$, who plays $c$ at every possible history. Short-run players hold a full-support prior belief $\mu_0 \in \Delta(\Omega)$ over the long-run player's type. 

For any strategy $\sigma$, define an associated belief system $\mu: H \to \Delta(\Omega)$, which outputs a belief about the long-run player's type at each $h^t$. Together, $(\sigma, \mu)$ is an \emph{assessment}. Given an assessment, let $\overline \sigma_L(h^t)$ denote the {unconditional} stage game strategy of the long-run player at $h^t$, averaging over $\mu(h^t)$, and $\sigma_L(h^t; \omega)$ be the stage game strategy of a committed type $\omega_c$ long-run player. Let $\sigma_L(h^t)$ be the rational long-run player's strategy. 
Analogously, let $\overline{\bb{P}}^\sigma$ denote the probability measure over histories generated by the unconditional strategy $\sigma$, and $\bb{P}^\sigma$ the measure generated when $\omega = \omega_R$. Finally, let $\sigma_L(\cdot | h^t)$ be the continuation strategy of the rational-type long-run player starting from some $h^t$. 

The main results of this paper study the set of (rational) long-run player payoffs that can be attained in sequential equilibria $(\sigma, \mu)$, defined as follows.

\begin{definition}
    An assessment $(\sigma, \mu)$ is a \emph{sequential equilibrium} if:
    \begin{enumerate}
        \item \und{Sequential Rationality}. At every history $h^t$, the continuation strategy $\sigma_L(\cdot | h^t)$ maximizes the long-run player's continuation payoff given $\{\sigma_i(h^s)\}_{i \in [N], h^s \succsim h^t}$ and $\mu(\cdot | h^t)$. 
        \item \und{Short-Run Nash Best Replies}. At every history $h^t$, $\{\sigma_i(h^t)\}_{i \in [N]} \in \mcal N(\overline \sigma_L(h^t))$. 
        \item \und{Bayesian Updating}. At every history $h^t \in \supp(\overline{\bb{P}}^\sigma)$, $\mu(h^t) = \mu(\cdot | h^t, \sigma)$, i.e. the conditional belief over $\Omega$ given history $h^t$ and strategy $\sigma$. 
        \item \und{Consistency.} There exists a sequence $(\sigma^n, \mu^n)$, with $(\sigma^n, \mu^n) \to (\sigma, \mu)$ such that\footnote{Because $\sigma: H \to \Delta(C)$ maps an infinite set into another, the notion of convergence is not immediately clear. Say $(\sigma^n, \mu^n)$ converges to $(\sigma, \mu)$ if $\sigma^n \to \sigma$ converges in the \emph{strong topology} (see Appendix B of \cite{Luo2025Paying}) and $\mu^n \to \mu$ converges weakly.} 
        \begin{enumerate}
            \item $\supp(\overline{\bb{P}}^{\sigma^n}) = H$ for all $n$, and $\mu(\cdot | h^t, \sigma^n) = \mu^n(h^t) \in \Delta(\Omega)$, and 
            \item $\sigma_L^n(\cdot, \omega_c) = c$ at every history $h^t$. 
        \end{enumerate}
    \end{enumerate}
    An assessment $(\sigma, \mu)$ is a \emph{Markov perfect equilibrium} if it is a sequential equilibrium and moreover at any histories $h^t, h^s$, 
    \[ \mu(h^t) = \mu(h^s) \implies \sigma(h^t) = \sigma(h^s) \]
\end{definition}

Consistency of equilibrium profiles will be useful when evaluating the value of deviations the long-run player may consider which are off the path of play (and not supported by some commitment type); absent some consistency-type condition, the value of such deviations would be hard to define in a principled way. Note the ``trembles'' defined here for consistency do not change the prior belief over $\Omega$ (which would then change $\mu^n$). A standard compactness argument shows this is without loss of generality (one could add the requirement $\mu_0^n \to \mu_0$ as well), but for simplicity I omit this additional notation.  

Markov perfection imposes the stronger assumption that at any two histories where short-run player posterior beliefs about long-run play are the same, players take the same actions.\footnote{It may seem unnatural to require the long-run player take the same action when short-run player beliefs are the same. The definition of Markov perfection can be relaxed to require only that short-run players strategies are measurable in $\mu$; however, since there is a single long-run player, their strategy can then be taken to be (without loss of generality) also measurable in $\mu$, less a public randomization device.}
While such a refinement may seem a-priori restrictive, I show it arises naturally when studying the effect exit has on reputational incentives (see Theorem \ref{t: commitment theorem}). 

Let $\und U$ and $\overline U$ be the long-run player's lowest and highest payoff from any sequential equilibrium in the game.\footnote{There is some potential ambiguity about whether sequential equilibrium need to exist when $C$ is infinite (see \cite{MyersonRenySequential} for some discussion and general fixes). However, given $\Omega$ is finite and $C$ is countable, the problem is substantially simpler---I show in Proposition \ref{p: folk theorem} and Theorem \ref{t: commitment theorem} that sequential equilibria as defined above exist, so in particular $\und U, \overline U$ are finite.}
Similarly, let $\und U^M$ and $\overline U^M$ be the long-run player's lowest and highest payoff from any MPE in the game.

The following assumption will be crucial to the analysis. 

\begin{definition}
    $(u, v, f)$ satisfies \emph{(weak) $c^*$ incentive compatibility} if for all $\beta \in [0, 1]$ and for each $\alpha \in \mcal N(\beta c^*)$, 
    \[ f(\alpha, c^*)[v(\alpha, c^*) + V_*] > (\geq) f(\alpha, 0)[v(\alpha, 0) + \overline{V_0}]\]
\end{definition}

\begin{assumption}
        \label{ass: c* IC}
Suppose that $(u, v, f)$ satisfies $c^*$ incentive compatibility. 
\end{assumption}

Assumption \ref{ass: c* IC} is a nontriviality condition that guarantees that, regardless of the initial belief $\mu_0(\omega_R)$, the action $c^*$ is enforceable: that is, it is an equilibrium of the game to play $c^*$ forever. It is clear that this is a necessary condition for there to be a (unique) Markov equilibrium in which $c^*$ is played forever; Theorem \ref{t: commitment theorem} shows that it is in fact also sufficient.

Note that Assumptions \ref{ass: dominance} and \ref{ass: c* IC} imply that the long-run player benefits from commitment; in particular, for the $\alpha$ that generates $\bar V_0$, we have that 
\[ f(\alpha, 0)[v(\alpha, 0) + V_*] > f(\alpha, c^*)[v(\alpha, c^*) + V_*] \geq f(\alpha, 0)[v(\alpha, 0) + \bar V_0] \]
first by Assumption \ref{ass: dominance} and then by Assumption \ref{ass: c* IC}. Hence $V_* > \bar V_0$. 

Throughout the analysis, I will maintain Assumptions \ref{ass: dominance} and \ref{ass: c* IC}. While some results will hold in more generality, it will be useful to maintain uniform assumptions throughout the exposition in order to compare the equilibrium payoff sets of the game as the reputation structure and the solution concept vary.

\section{Commitment Payoff Theorems}

This section contains the main results of this paper, which characterize the payoff set in the game with and without reputation, in sequential and Markovian equilibria. Informally, I establish the following table (supposing Assumptions \ref{ass: dominance} and \ref{ass: c* IC} hold), which summarizes the takeaway from the results in this section.

\begin{table}[h]
\begin{center}
    \begin{tabular}{|c|c|c|}
 \hline              & non-Markov      & Markov                   \\ \hline
Reputation    & Anti-Commitment Theorem (Prop. \ref{p: exit hurts reputation}) & Stackelberg Payoff (Theorem \ref{t: commitment theorem})       \\ \hline
No Reputation & Folk Theorem (Prop. \ref{p: folk theorem})    & Static Nash Payoff (Prop. \ref{p: markov folk theorem}) \\ \hline
\end{tabular}
\end{center}
\end{table}

\vspace{-2em}

\subsection{Complete Information}
We start by establishing some simple facts about the complete information game where there is no uncertainty about the long-run player's rationality. Say a \emph{folk theorem} holds in the game if the set of payoffs attainable as part of some equilibrium in the complete-information game contains the set of feasible and individually rational payoffs for the long-run player (here the set $[\und V_0, V^*]$), supposing short-run players myopically best reply. 
Proposition \ref{p: folk theorem} shows that a folk theorem holds in our game. 
Arguments from \cite{fudenberg1994efficiency}, \cite{fudenberg1994folk} do not necessarily immediately apply because exit can ``loom large,'' and thus lead to periods where the effective discount rate is small even without discounting, i.e. $\delta = 1$. Despite this, we have the following result. 

\begin{proposition}
\label{p: folk theorem}
Let $\mu_0(\omega_R) = 1$. If $\bar c$-IC holds, then 
$[\und V_0, V^*] \subset [\und U, \overline U]$. 
\end{proposition}

The proof is constructive. First, $\und V_0$ is the worst equilibrium payoff since short-run players must myopically best reply. That the long-run player can do better follows from constructing an equilibrium where the long-run player must always play their upper Stackelberg action, leading to myopic best replies; Assumption \ref{ass: c* IC} ensures that this is enforceable as an equilibrium profile. 

Importantly, though, the equilibria constructed in the proof of Proposition \ref{p: folk theorem} to sustain a high payoff for the long-run player require that punishments off the equilibrium path are non-Markov, in the sense that they are not measurable in $\mu_t(h^t)$ (which is constant in the complete information case). Given this observation, it is perhaps unsurprising that the Markov restriction can strongly refine the set of payoffs: in the complete information game, it does so exactly by holding the long-run player down to their static payoff. 

\begin{proposition}
\label{p: markov folk theorem}
   Let $\mu_0(\omega_R) = 1$. Then $[\und U^M, \overline U^M] = [\und V_0, \overline V_0]$.     
\end{proposition}

This property of the complete-information Markov game will be important in disciplining the value of deviations in the game with reputational concerns. The proof follows immediately once it becomes clear that the Markov refinement requires the continuation strategies to be independent of the history (since $\mu(h^t) = 1$ at all histories in all sequential equilibria). For completeness, I give a formal argument in the \blue{\ref{markov folk theorem proof}} Note that Proposition \ref{p: markov folk theorem} applies to any repeated game with a dominant long-run action, not necessarily just those with exit.

\subsection{Low Payoffs and Sequential Equilibrium}
Before stating the main positive reputation selection result---which is stated in terms of Markov strategies---I first consider reputation effects in general sequential equilibrium. 
Without putting additional restrictions on the survival probability function $f$, is it possible to say anything about when the long-run player can establish a reputation outside of Markov strategies?
In this subsection, I show the answer is no, even when the ``effective'' discount rate is high enough that the Stackelberg action is enforceable. Formally,

\begin{proposition}[Anti-Commitment]
    \label{p: exit hurts reputation}
    Fix $(u, v)$ such that 
    \begin{enumerate}
        \item $\{0\} = \argmax_{c \in C} v(a, c)$ for all $a$, and 
        \item For some $\hat c$, $\overline{\U_{\alpha \in \mcal N(0)} \supp(\alpha)} \cap \overline{\U_{\alpha \in \mcal N(\hat c)} \supp(\alpha)} = \varnothing$. 
    \end{enumerate}
    Then for all $M, \varepsilon > 0$, there exists $\xi > 0, f$ satisfying Assumptions \ref{ass: dominance} and \ref{ass: c* IC} such that 
    \begin{enumerate}
        \item $\bar U \geq V_* \geq M$. 
        \item $\limsup_{\mu_0(\omega_R) \to 1} \und U  \leq \varepsilon$. 
    \end{enumerate}
\end{proposition}

Proposition \ref{p: exit hurts reputation} should be interpreted as follows. As $\mu_0(\omega_R) \to 1$, the set of sequential equilibrium payoffs with reputation grows arbitrarily large, in stark contrast to the baseline payoff selection results of \cite{FudenbergLevine89} without exit. This is stronger than the simple statement the long-run player's payoff can be arbitrarily low, and highlights the way that endogenous exit changes incentives in ways that are more subtle than mechanically decreasing the discount rate.\footnote{This simpler statement can be attained by setting $f(a, c)$ to be arbitrarily low for all $(a, c)$.}
In particular, Part (1) of Proposition \ref{p: exit hurts reputation} guarantees that (a) the endogenously chosen discount rate is still large enough that the commitment action is enforceable, and (b) that repeatedly playing the commitment action gives the long-run player a high payoff. 
Despite this, Part (2) of Proposition \ref{p: exit hurts reputation} states that, even in the presence of reputational incentives, the long-run player's payoff can be arbitrarily low.
Indeed, Proposition \ref{p: exit hurts reputation} is a particularly strong indeterminacy result: it says for any payoffs $(u, v)$ satisfying our dominance condition one can find an exit probability function such that the set of equilibrium payoffs as $\mu_0(\omega_R) \to 1$ is at least $[\varepsilon, M]$ for any $\varepsilon, M \geq 0$. 

Condition (1) is a standard dominance condition, and is a necessary feature of the primitives for the payoffs to inherit the dominance condition of Assumption \ref{ass: dominance}. Condition (2) is a disjointness assumption that is required so that $f$ can be consistently well-defined. To see why it is necessary, suppose that $\mcal N(0) = A$. Then there is no way to construct the continuation probabilities in a way that is continuous and also varies with the worst-case selection. It turns out condition (2) is the right minimal condition given the construction (which I describe below) that delivers the result. Note Condition (2) is satisfied by both the examples in Sections 4 and 5. 

The construction proceeds roughly as follows. First, set $f(a, c) = f(a)$ to be independent of $c$, so that Assumption \ref{ass: dominance} is inherited by our restriction on $v$. Second, for every $a^* \in \mcal N(c^*)$, set $f(a^*) = 1 - \xi$. This guarantees that $V_* \geq \min_{a^* \in \mcal N(c^*)} \frac{v(a^*, c^*)(1 - \xi)}{\xi}$, which grows larger than $M$ as $\xi$ grows small. This then in turn guarantees $c^*$-incentive compatibility regardless of the action $a$, so $c^*$ is always enforceable and Assumption \ref{ass: c* IC} is satisfied. Proposition \ref{p: folk theorem} then implies the first part of the statement, since on the pooling path both types play $c^*$, so short-run best replies are the same as under complete information and the profile in Proposition \ref{p: folk theorem} remains a sequential equilibrium of the perturbed game. For the second part, fix any $\und a \in \mcal N(0)$ and set $f(\und a) = \eta$ for some $\eta$ small enough. Construct an equilibrium that rewards $V_*$ regardless of continuation play in the first period, noting this then means playing $0$ is a best reply in period $1$ by Assumption \ref{ass: dominance}. Then there exists an equilibrium where $f(\und a)[v(\und a, 0) + V_*] \leq \eta[v(\und a, 0) + V_*]$. Taking $\eta \to 0$ for fixed $M$ then implies the result. 
Explicit computations are collected in the \blue{\ref{p: exit hurts reputation proof}}

\subsection{Guaranteeing Stackelberg Payoffs}
Having shown that reputation effects no longer select payoffs in general sequential equilibria, a natural next question is whether there is a refined class of equilibria under which one can obtain a positive payoff selection result in the game with exit. 

Recall that classical reputation results hold only after many\footnote{``Many'' here depends formally on the speed of learning and the prior belief of rationality; see \cite{Gossner11} for a more formal discussion.} (finite) periods have passed, as the long-run player must teach their opponents that they are playing the commitment action. Yet in the game with exit, they may not have the luxury of time; the game may have ended far before short-run player beliefs about long-run play have concentrated on $c^*$, keeping the long-run player from securing a high payoff. 
Thus exit perturbs the long-run player's strategic incentives in two ways: first, by making the future less valuable (and hence the promise of future payoffs less motivating to discipline signaling behavior today), and second, by shortening the length of time they have to convince short-run player beliefs about their (committed) play. 

The first problem, as evidenced by Proposition \ref{p: folk theorem}, can be resolved by adding an appropriate incentive compatibility constraint for the long-run player. The second problem, however, is not so easy to fix: indeed, I will need the full force of a Markovian refinement in order to iron out beliefs which are adversarial to guaranteeing the long-run player their Stackelberg payoff. The technical contribution of Theorem \ref{t: commitment theorem} is that these two ingredients essentially characterize when one can obtain a commitment payoff theorem in the game with exit. 

\begin{definition}
    Primitives $(u, v, f)$ are \emph{monotone} if $A$ can be ordered such that for each fixed $c$, $f(a, c)$ is strictly decreasing in $A$, $v(a, c)$ is nonincreasing in $A$, and $\mcal N(\beta c)$ is decreasing (in the strong set order) in $\beta$.
\end{definition}

\begin{theorem}[Markov Commitment]
\label{t: commitment theorem}
Suppose $\mu_0(\omega_R) < 1$. In all Markov equilibria $\sigma$, $\sigma_L(h^t) \equiv c^*$ for all $h^t \in \supp(\bb{P}^\sigma)$. In particular, $\und U^M = V_*$. 

   Conversely, if $(u, v, f)$ is monotone, $\mcal N(0)$ is single-valued, and the long-run player does not satisfy weak $c^*$ incentive compatibility, then 
   $ \limsup_{\mu_0(\omega_R) \to 1} |\und U^M - V_*| > 0$.
\end{theorem}

Theorem \ref{t: commitment theorem} is the main theoretical takeaway of this paper. First, it identifies a sufficient condition---$c^*$-incentive compatibility---that negates the effect of exit and ensures the long-run player their Stackelberg payoff in Markov equilibrium. Second, it shows (relative to Proposition \ref{p: markov folk theorem}) the long-run player attains a strictly {higher} payoff when they have incentives for reputation formation than their best Markov payoff in the game without reputation. Third, under a monotonicity condition, failure of weak $c^*$ incentive compatibility prevents the lower Markov payoff from converging to the Stackelberg payoff as $\mu_0(\omega_R) \to 1$. Fourth, it establishes existence of Markov equilibrium, which is not a-priori immediate from the definitions.

Several features of Theorem \ref{t: commitment theorem} are worth discussing. First is the role of $c^*$-incentive compatibility, which ensures both that $c^*$ is enforceable (when $\beta = 1$) and that it remains enforceable when short-run players best respond to a mixture of $0$ and $c^*$. The latter is critical to discipline long-run player incentives and ensure mixed-strategy Markov equilibria do not arise on the path of play (see Lemma \ref{l: markov no mixing}). 

The second is the role of the Markov refinement, which is best understood in comparison to Proposition \ref{p: exit hurts reputation}. In particular, the Markov refinement disciplines the feasible payoff set after rationality is revealed (thus decreasing incentives for ``revealing rationality''), making it hard for short-run players to reward in the future actions which are non-Stackelberg. 
It also ensures short-run players believe the commitment action will be played whenever reputation foundations have bite, which ensures that exit does not  ``loom large'' in the first period.
Markov beliefs thus rule out equilibria where the long-run player is rewarded in the future in a complicated way for revealing they are rational today, even though doing so will cause the game to end with arbitrarily high probability today.  

The Markov refinement can thus be more broadly interpreted as follows. In general reputation games, there are two forces at play: a backwards looking one (is past long-run player behavior consistent with play by the commitment type?) and a forward looking one (does this imply future long-run player behavior will be consistent with Stackelberg play?) In traditional reputation models, these two forces together ensure the long-run player can eventually teach short-run players their action is indistinguishable from that of the committed type regardless of the equilibria, guaranteeing a high payoff bound. The Markov refinement preserves the backwards looking force (learning starts at $t = 0$), but cannot preserve the forward looking one since exit might occur before ``future long-run player behavior'' matters. Consequently, the Markov refinement shows that in games with exit, measurability of play in beliefs is a substitute to the forward-looking force, and preserves a payoff selection theorem. 

Finally, the converse. Monotonicity of the game is satisfied in natural applications, including the coordination game later studied in this paper. For example, it is satisfied whenever the exit probability increases with ``more aggressive' actions $a \in A$, $v$ depends only on the long-run player's action, and short-run players play a strictly submodular game with unique best replies (as is the case in the globalized coordination game analyzed later) in $(a, c)$ so that higher levels of long-run player actions lead to less aggressive sets of actions. Monotonicity is sufficient to imply that when the long-run player is indifferent between several actions, they cannot secure a higher payoff than their pure Stackelberg payoff. This rules out situations where the long-run player benefits from mixed Nash equilibrium. In particular, it ensures that when mixing occurs in the first period (which is used to prove the converse to Theorem \ref{t: commitment theorem}), the long-run player must strictly underperform their commitment payoff in such an equilibrium. 
Note the converse is an additional, independent contribution and relies on the exit structure. In particular, it implies that the Stackelberg upper bound result is tight in the modeling assumptions, which is different from standard reputation arguments, where the long-run player may attain the Stackelberg payoff even when $\delta$ is far from $1$. 

I conclude this section with an outline of the proof of Theorem \ref{t: commitment theorem}. 
\begin{proof}
The forward direction. Suppose the long-run player satisfies $c^*$ incentive compatibility. We show in all Markov perfect equilibria the long-run player plays their commitment action $c^*$ and short-run players best respond, securing the long-run player $V_*$. 
To do so, we establish a number of lemmas whose (technical) proofs are relegated to the \blue{\ref{sequential learning proof}}

First, we must discipline the inferences and payoffs that short-run players have when the long-run player does not imitate the commitment type. 
\begin{lemma}
\label{l: sequential learning}
    Let $(\sigma, \mu)$ be a sequential equilibrium. For all $h^t \neq \{c^*\}_{s = 0}^{t - 1}$, $\mu(h^t)(\omega_R) = 1$. 
\end{lemma}

Lemma \ref{l: sequential learning} (along with Proposition \ref{p: markov folk theorem}) guarantees that, off the equilibrium path, the highest payoff the long-run player can obtain is $\overline V_0$.
Our next lemma disciplines the set of actions the long-run player can take in any equilibrium. 

\begin{lemma}
\label{l: markov two actions}
   Let $(\sigma, \mu)$ be Markov perfect. Then for all $h^t$, $\supp(\sigma_L(h^t)) \subset \{0, c^*\}$. 
\end{lemma}

Lemma \ref{l: markov two actions}, along with Lemma \ref{l: sequential learning} are enough to imply (following an argument similar to the first part of the proof of Proposition \ref{p: folk theorem} and taking $\beta = 1$), that always playing $c^*$ {is} a Markov perfect equilibrium. This gives existence of an MPE. 

Our last step is to show that this is the only behavior for the long-run player consistent with Markov perfect equilibrium. Clearly, always playing $0$ cannot be Markov perfect under assumption. To see why, take $\beta = 1 - \mu_0(\omega_R)$ and fix any $a \in \mcal N(\beta c^*)$. Rearranging the incentive compatibility constraint implies 
\[ f(a, 0)[v(a, 0) + \overline V_0] < f(a, c^*)[v(a, c^*) +  V_*] \]
so the long-run player can strictly profitably deviate by always choosing $c^*$ in any period they were supposed to deterministically play $0$ where $\mu(h^t)(\omega_R) < 1$, e.g. the first period. 
Finally, we show that the long-run player prefers not to mix in Markov equilibria. 
\begin{lemma}
\label{l: markov no mixing}
   Let $(\sigma, \mu)$ be Markov perfect. Then for all $h^t$, $|\supp(\sigma_L(h^t))| = 1$. 
\end{lemma}

The intuition is that in order to sustain mixing, the long-run player must be indifferent between their upper Nash and lower Stackelberg payoffs, modulo the cost today. When this difference is large and exit does not shut the game down early, then this can never be case, which ensures the long-run player is never indifferent between actions. 

The proof of the converse is constructive and can be found in the \blue{\ref{converse commitment theorem proof}} The basic idea is that, at any $\beta$ where the incentive compatibility constraint holds with equality, we find an equilibrium where the long-run player mixes at the time $t = 0$ public history. Monotonicity of the game then ensures that at mixing histories, they underperform their commitment payoff. The condition that $\mu_0$ is small rules out equilibria where $\mu_0(\omega_{c^*}) \approx 1$ and the long-run player can secure a very high payoff in the first period even if they play $0$. 
\end{proof}

\section{The Chain Store Game}

Consider the canonical chain store game, with two variations: first, suppose the game is infinitely repeated, so the monopolist must continually face entrants over and over again, and second, suppose player actions can lead to exit in the game. The first assumptions implies that, without exit, the results of \cite{FudenbergLevine89} apply, and hence the long-run player can secure their Stackelberg payoff in all Nash equilibria as they become extremely patient. The goal of this section is to understand how exit perturbs these results. In incumbent-entrant games, exit may be a particularly appropriate assumption: accommodating an entrant can lead to the incumbent being displaced from their position, while fighting may increase the probability they remain an incumbent even after entry. 
This can lead to an additional channel for and challenges to reputation formation. 

Formally, suppose in each period the long-run player ($I$) chooses to fight or Accommodate an entrant, but only after the entrant ($E$) has decided whether or not to enter.  
Let stage game payoffs be as follows, where the first entry is the long-run player's payoff. Let $\delta \in (0, 1)$. \footnote{These specific payoffs for the chain store game are due to \href{https://bpb-us-e1.wpmucdn.com/sites.northwestern.edu/dist/4/2519/files/2023/04/Lecture_1.pdf}{Harry Pei's Lecture Notes.}}
Interpret here ``Accommodate'' as the action $0$. 

\begin{table}[ht]
    \centering
    \begin{tabular}{|c|c|c|}
    \hline 
     & Fight & Accommodate \\ \hline
        In  & $(0,-1)$        & $(2,1)$ \\ \hline 
        Out & $(2-\delta,0)$ & $(2,0)$ \\ \hline 
    \end{tabular}
    \caption{Payoffs in the Chain Store Game}
    \label{tab:chain-store-payoffs}
\end{table}




Moreover, define $f(a, c)$ to take the values in Table \ref{t: table 1}, where $\varepsilon, \eta$ are strictly positive and $1 - \eta \geq \frac23$. 

\begin{table}[h]
\centering
\begin{tabular}{|c|c|c|}
\hline
    & Fight      & Accommodate    \\ \hline
In  & $\varepsilon$  & $\varepsilon$ \\ \hline
Out & $1 - \eta$ & $1 - \frac{\eta}{2}$    \\ \hline
\end{tabular}
\caption{Survival Probabilities for the Chain Store Game}
\label{t: table 1}
\end{table}

Note that these payoffs imply Accommodate is dominant.
The chain store game admits the following result. 

\begin{proposition}
    \label{p: chain store game}
  Let 
\[ K = \frac{(1 - \eta)(2 - \delta)}{\eta}, \quad \varepsilon_s = \frac{K}{K + 2}, \quad \bar \varepsilon = 1 - \frac{2}{K}.\] Then for all $\varepsilon \in (0, \varepsilon_s)$, $F$ is the commitment action and the game is monotone. Moreover, 
\begin{enumerate}
    \item If $\varepsilon \in (0, \bar \varepsilon)$, $U^M = V^* = K$. 
    \item If $\varepsilon \in (\bar \varepsilon, \varepsilon_s)$, weak $F$-incentive compatibility fails, and so 
    \[ \limsup_{\mu_0(\omega_R) \to 1} |U^M - V_*| > 0.\]
    \item For all $\varepsilon \in (0, \bar \varepsilon)$, 
    \[ \limsup_{\mu_0(\omega_R) \to 1} \und U \leq \varepsilon (2 + K). \]
\end{enumerate}
\end{proposition}

The formal proof is in the \blue{\ref{p: chain store game proof}} The intuition applies Theorem \ref{t: commitment theorem} to show that as $\varepsilon$ vanishes, payoffs satisfy Assumption \ref{ass: c* IC}; since the game is monotone in the order $In \succ Out$, this implies the (and only) part of the result. The last part follows by taking $\varepsilon$ small and constructing an equilibrium as in the proof of Proposition \ref{p: exit hurts reputation}. 

Three observations are in order. First is that in Proposition \ref{p: chain store game}, the key parameter was the survival probability given that the entrant entered the game. Intuitively, varying $\varepsilon$ varies the effective discount rate but also the incentives to take different actions; lowering $\varepsilon$ increases the ``cost'' of Nash equilibrium and makes reputation formation more attractive to the long-run player, especially relative to fighting. 
Second is the contrast of Proposition \ref{p: chain store game} with known results from reputation a la \cite{FudenbergLevine89}. In particular, the third part of the proposition highlights how reputation can fail with exit even with extremely patient players who have no time discounting. This shows that the Markovian refinement is critical for any hope of selecting a unique equilibrium, exactly because it rules out nonstationary equilibrium where the exit cost becomes ``sunk'' to the long-run player, who chooses their stage game action conditional on the fact they have already survived today. Finally, I am able to give a unique behavioral prediction in Markov strategies in this game in which the incumbent always fights and short-run players never enter (i.e. deterrence is successful), unlike the indeterminacy of behavioral predictions in reputation games in general (c.f. \cite{PeiLi21}).

\section{Regime Change}
\subsection{The Global Games Model}
Consider the problem faced by an authoritarian government (player $L$) who must decide how they should punish dissidents who engage in a revolution. In each period, having successfully defended against an attack, the government can choose to punish dissenters by decreasing their per-period utility by $c$ at a cost of $\kappa(c)$, where $\kappa$ is a continuous strictly increasing function satisfying $\kappa(0) = 0$. Let $C = [0, \overline c]$ and moreover suppose $\kappa(\overline c) > 1$, so that the government never chooses the maximally feasible value of punishment.\footnote{Note $C = [0, \overline c]$ is distinct from the assumption made in the main model that $C$ is countable. This is useful in defining $c^*(0)$, after which it is sufficient to assume $C = \{0, c^*(0)\}$ for the remainder of the application.} \footnote{This is an Inada-type condition that ensures the government's problem is always well-defined. Throughout the analysis, we restrict to $\{c : \kappa(c) < 1\}$.}

The (unit mass, continuum) dissidents now face the following problem. In each period, they know the government is able to fend off $\theta$ attackers, where $\theta \sim P_\theta$ is drawn independently in each period, with $\bb{E}_{P_\theta}[\theta] = \mu$ and $\var(\theta) = \tau$.\footnote{In practice, I will consider the case where $\tau$ is particularly small. This can be interpreted as a world where the regime has strength $\mu$ and experiences a small idiosyncratic shock in each period to that strength.} Each dissident $i$ cannot see $\theta$ perfectly, instead observing an imperfect signal $x_i$ of $\theta$ with some idiosyncratic noise, $x_i = \theta + \varsigma \varepsilon_i$, where $\varepsilon_i \sim F_\varepsilon$ is i.i.d. across each player. Suppose players' information sets are sufficiently well-behaved that they satisfy the following regularity conditions:
\begin{assumption}[Informational Regularity]
\label{ass: information structure} The following hold. 
    \begin{enumerate}
        \item $F_\varepsilon$ and $P_\theta$ admit strictly positive, bounded continuous densities; $F_\varepsilon$ is symmetric around $0$, and $P_\theta$ is symmetric around $\mu \in (0, 1)$. 
        \item The conditional distribution $F_{\theta}(\cdot | x)$ has strictly positive differentiable density and is decreasing in its second argument.
        \item The conditional distribution $F_x(\cdot | x)$ has strictly positive differentiable density and is decreasing in its second argument. 
    \end{enumerate}
\end{assumption}
Here, $F_\theta(\cdot | x)$ is the first order belief over fundamentals of players supposing they see signal $x$ and have prior $P_\theta$; $F_x(\cdot | x)$ is the second-order conditional belief about other players' signals (i.e. player $i$'s thought about others' signals after receiving $x$). The symmetric assumption on $F_\varepsilon$ allows it to be interpreted as white noise. I require $\mu \in (0, 1)$ so that the average state is not in a dominance region. 

Assumption \ref{ass: information structure} also guarantees $F_x$ admits a strictly positive density and that $F_x(x | \theta)$ is decreasing in the realized state; moreover, $F_\theta(\cdot | x)$ and $F_x(\cdot | x)$ are differentiable as the noise term is additive. These smoothness assumptions ensure a first order approach for players is valid, and generalize known conditions standard in the global games literature.\footnote{I am not aware of a global games setting studied in the literature that does not satisfy these assumptions; for example it nests normal-normal information structures.}

Having seen their private signals, the dissidents independently and simultaneously choose an action $a_i \in \{0, 1\}$ in each period, with $a = A = \int a_i di$ equal to the total mass of agents taking the aggressive action (i.e. revolt), $a_i = 1$. 
If enough dissidents attack (i.e. $A \geq \theta$), then the revolt is successful, the regime is overthrown, and the game ends. The continuation probability $f(A, c)$ is exactly the probability the revolt fails, is given by $f(A, c) = \bb{P}_\theta(A \leq \theta)$, and is independent of $c$. 

Players' payoffs are given by 
\[   v(A, \theta, c) =  \begin{cases} 0 & \text{   if  } \theta < A \\ 1 - \kappa(c) & \text{  if  } \theta \geq A \end{cases} \quad  \text{  and  } \quad  u_i(a_i, A, \theta, c) = \begin{cases} a_i & \text{  if   } \theta < A \\ - a_i c & \text{  if   } \theta \geq A \end{cases}. \]
Note here payoffs mirror standard global games payoffs (see i.e. \cite{MorrisPauznerFrankel03}) where $\theta$ is the peg of the regime, but with the cost of failed coordination (i.e. the coordination friction) flexibly set by the regime at some cost. 
Standard arguments from the global games literature imply the following. 

\begin{proposition}
\label{p: stage game properties}
  Suppose $\varsigma < \tilde \varsigma$ for some $\tilde \varsigma > 0$ sufficiently small. Then for all $c$, there exists functions $x^*, \theta^*: \bb{R}_{++} \to \bb{R}$ such that 
  \begin{enumerate}
      \item The strategy $s(x, c) = \mathbf{1}\left\{x < x^*(c) \right\}$ is the (essentially) unique strategy surviving iterative deletion of strictly dominated strategies. 
      \item $\theta^*(c)$, is strictly decreasing in $c$, and $\int s(x, c) dF_{x | \theta} \geq \theta$ if and only if $\theta \leq \theta^*(c)$. 
  \end{enumerate}
\end{proposition}

Proposition \ref{p: stage game properties} implies not only that $\mcal N(c)$ is single-valued, but that both the set of states which successfully guard against attacks and the cutoff state are themselves monotone at a cutoff, whose evolution is well-behaved. These properties will be useful in the proof of Theorem \ref{t: global games theorem} and is proven in the \blue{\ref{p: stage game properties proof}}
Note Assumption \ref{ass: dominance} is satisfied as well as $1 - \kappa(c) < 1 - \kappa(0)$ for all $c > 0$, and the exit probability depends only on $A$. 

Finally, the game is clearly monotone when $A$ is ordered in the standard way in $\bb{R}$. Hence the globalized game satisfies the assumptions of the general model. 

\subsection{Selecting Equilibria}
When is it that the threat of punishment is enough to deter revolution? Since the game is monotone, $c^*$-incentive compatibility is both necessary and sufficient for the regime to deter revolution (as much as possible) in the unique Markov equilibrium. 
Rewriting the incentive constraints to the simpler setting of the global game described above yields that $c^*$-incentive compatibility is equivalent to the inequality
 \[ V_* - \und V_0 > \kappa(c^*).\]
 In the global games setting, $c^*$ incentive compatibility is related to the degree of dispersion of higher order beliefs around the risk-dominant threshold {at the average}, i.e. whether $\mu > (<) \frac{1}{1 + c^*}$. Here, recall value $\frac{1}{1 + c^*}$ is the risk-dominant signal threshold selected by the global game as signals become arbitrarily precise, i.e. $\lims_{\varsigma \to 0} x^*(c^*) =  \frac{1}{1 + c^*}$ by \cite{MorrisPauznerFrankel03}.
 
 What is the right commitment posture $c^*$ to consider when comparing to $\mu$? There are two concerns. First, since $c^*(\varsigma)$, the optimal commitment action as a function of $\varsigma$, is nonconstant, the risk-dominant selection threshold varies with the noise precision as well. Second, the underlying probability distributions $(F_\varepsilon, P_\theta)$ will also affect the optimal choice for the regime and affect their commitment payoff. To resolve this problem, fix $(F_\varepsilon, P_\theta)$ and consider the set 
 \[ c^*(0) = \limsup_{\varsigma \to 0} c^*(\varsigma)  \]
 where $\limsup A_n$ are the subsequential limits of all selections from $A_n$. That $c^*(0)$ is nonempty follows from a compactness argument, noting $c^*(\varsigma)$ is well-defined and nonempty for each $\varsigma$ and hence so must $c^*(0)$. 
 Throughout, I will fix some $c^* \in c^*(0)$ and appeal to continuity of the problem in parameters to make the argument that the same statement holds for $c^*(\varsigma)$.

Throughout the statement of Theorem \ref{t: global games theorem}, suppose $\Omega = \{\omega_R, \omega_{c^*}\}$ for some $c^* \in c^*(0)$. I also assume by fiat that $c^* \neq 0$, which is assumed throughout the main exposition, so that the commitment type is distinguishable from one playing static Nash.\footnote{This assumption can be derived by putting mild regularity conditions on $\kappa$, for example that $\kappa'(0) = 0$.}

\begin{theorem}
    \label{t: global games theorem}
   For all $\eta > 0$, there exists $\overline \tau > 0$ and $\overline \varsigma(\tau)$ vanishing as $\tau \to 0$, such that for $(F_\varepsilon, P_\theta)$ satisfying $\var(\theta) = \tau < \overline \tau$, and $\varsigma < \overline \varsigma(\tau)$, then 
   \begin{enumerate}
       \item If $\frac{1}{1 + c^*} < \mu$,  $\und U^M = V^*$ and $f(a^*(c^*)) > 1 - \eta$.  
       \item If instead $\frac{1}{1 + c^*} > \mu$, then $\max\{\und U, \overline U^M\} < \eta$. 
   \end{enumerate}
\end{theorem}

Theorem \ref{t: global games theorem} highlights a stark difference between the regime's Stackelberg payoff depending on where their {average} regime strength when (1) the average becomes an arbitrarily precise summary of the state in each period and (2) the global games perturbation vanishes. Importantly, it delivers a novel implication of risk dominance that is absent from the static game: the regime can only successfully deter rebellion if and only if they are, on average, able to deter rebellion at the risk-dominant level. Should they not be able to do this, then exit immediately happens with high probability. 

The result stands in contrast to two seminal papers in the literature on dynamic global games with exit. First, it augments the model of \cite{AngeletosHellwigPavan07} to (1) small i.i.d. shocks to the state $\mu$ and (2) reputational incentives for the regime and shows how these perturbations can select a version of the ``survival dominant'' equilibrium (i.e. survival only if $\mu > \frac{1}{1 + c^*}$) as the unique Markov equilibrium. Finally it augments the reputational global game model of \cite{Huang17} to allow for small i.i.d. shocks and a rich exit process, and shows how the regime can fail if their average strength is too weak. This stands in stark contrast to his main result, where there are no attacks that ever occur on the equilibrium path. The differences in these results suggest that taking exit incentives seriously in reputational models can lead to rich results which are more indicative of real-world phenomena, where successful attacks do happen if speculators sense the regime is ``likely weak.'' 

The intuition behind the forward direction of Theorem \ref{t: global games theorem} follows from Theorem \ref{t: commitment theorem} and a standard concentration inequality argument: if $\mu > \frac{1}{1 + c^*}$, then the cutoff state as the noise vanishes is below the mean, so as the variance vanishes the regime must survive most of the time. Conversely, if $\mu < \frac{1}{1 + c^*}$, then most of the time the regime dies, and hence they must attain a low payoff. A formal proof can be found in the \blue{\ref{t: global games theorem proof}}

\section{Discussion}
This paper studied a model of reputation where players' actions endogenously affected the probability of exit in the game. Such endogeneity creates several difficulties for the analysis of the repeated game. First, folk theorems need no longer hold because the future can be ``small'' relative to actions today, making enforceability infeasible. Second, the addition of reputation effects need not select the long-run player's Stackelberg payoff even when the long-run player becomes extremely patient. 

I identify two conditions that help ameliorate these problems. These conditions are an \emph{incentive compatibility} condition on the effective rate of future discounting: the long-run player wishes to play the Stackelberg action if recommended to do so, supposing maximal variation in continuation payoffs after; and a \emph{Markov belief} condition, represented by the Markov refinement: short-run players believe the long-run player has an incentive to build a reputation when those incentives are present.
Applied to a global games setting, these conditions yield new predictions about the survival of equilibria relative to the risk dominant threshold in a dynamic regime change game. 

There are several fruitful directions for future analysis. First, understanding the limits of endogenous exit when players' actions are imperfectly monitored is a natural next step. Moreover, the paper focuses only on pure Stackelberg strategies. Accommodating mixed Stackelberg strategies and imperfect monitoring is natural and potentially interesting. The methods developed in this paper need not apply because reputation formation and incentive provision are much more subtle with noisy monitoring, and the standard law-of-large-numbers intuitions do not apply with endogenous exit. Second, understanding the case with multiple commitment types is also promising, away from sequential equilibria---here, the set of payoffs may be pathological because of continuation payoffs at off-path deviations (i.e. taking a committed action that is not on the path of play). Third, there is scope to apply the machinery in this paper to several other settings. For example, in the repeated trust game, my findings can give predictions about when the long-run player can teach short-run players to trust them even when a lack of trust might prematurely end the game. Alternatively, Theorem \ref{t: commitment theorem} can be used to give conditions under which Markov equilibria have reputation foundations in stopping time games where short-run or myopic players can stop at any time (i.e. adapted versions of \cite{Chassang10}). Finally, outside of Markov predictions, these results may give new interpretations of ``bad reputation'' style results in the spirit of \cite{ElyValimaki03} and \cite{ely2008when}. 

\bibliography{biblio.bib}

\appendix
\section*{APPENDIX: OMITTED PROOFS}
\makeatletter\def\@currentlabel{Appendix.}\makeatother
\label{Appendix A}

\subsection*{PROOF OF PROPOSITION \ref{p: folk theorem}}
\label{folk theorem proof}
\begin{proof}
Some preliminaries. Since $\mu_0(\omega_R) = 1$, we consider assessments where $\mu(h^t)(\omega_R) = 1$ always and find sequential equilibrium.
Finally, throughout the proof, assume players have access to a public randomization device so that the payoff set is convex. 

It is clear that $\und V_0$ is attainable by an equilibrium where the lowest stage game Nash equilibrium is played at every history. 
To attain $V^*$, let $\bar c$ be an upper Stackelberg action with $\bar a \in \mcal N(\bar c)$ together attaining $V^*$. Consider the following strategy profile. 
\begin{enumerate}
    \item If $h^t = \{\bar c\}_{s = 0}^{t - 1}$, $\sigma_L(h^t) = \bar c$, $\{\sigma_i(h^t)\}_{i \in [N]} = \bar a \in \mcal N(\bar c)$.
    \item At any other history, $\sigma_L^*(h^t) = 0$ and $\{\sigma_i(h^t)\}_{i \in [N]} = \und a \in \mcal N(0)$, with $\und a$ chosen to attain the lowest Nash equilibrium payoff $\und V_0$. 
\end{enumerate}
Clearly short-run players are myopically best replying. At off-path histories, the long-run player is also best replying: short-run player is chosen independently of their action, so by Assumption \ref{ass: dominance} playing $0$ is optimal. Finally, enforceability at $\bar c$ follows by noting that 
\[ f(\bar a, \bar c)[v(\bar a, \bar c) + V^*] \geq f(\bar a, 0)[v(\bar a, 0) + \und V_0] \]
by $\bar c$-incentive compatibility applied at $\beta = 1$. 
\end{proof}

\subsection*{PROOF OF PROPOSITION \ref{p: markov folk theorem}}
\label{markov folk theorem proof}
\begin{proof}
First, note that if $\mu_0(\omega_R) = 1$, then it must be that in every sequential equilibrium $\mu(h^t)(\omega_R) = 1$ as well for all on-path histories (as the agent has a degenerate prior belief). 
Now fix any history $h^t$; at any successor history $h^s, h^{s'} \succ h^t$, it must be that $\sigma(h^s) = \sigma(h^{s'})$, so in particular future play cannot vary based on play today. Thus, the long-run player's action must be myopically optimal, since 
\[ 0 \in \argmax_{c \in C} f(\alpha, c)[v(\alpha, c) + m]\]
for all $\alpha$, and hence $\sigma_L(h^t) = 0$ at any history $h^t$. Short-run players must play the {same} myopic best response at each history, so in equilibrium $\{\sigma_i(h^t)\}_{i \in [N]} \in \mcal N(0)$ always. Hence all equilibria are repeated static Nash. Public randomization convexifies the payoff set. 
\end{proof}

\subsection*{PROOF OF PROPOSITION \ref{p: exit hurts reputation}}
\label{p: exit hurts reputation proof}
\begin{proof}
    First, note that if condition (2) is satisfied, so that 
    \[ \overline{\U_{\alpha \in \mcal N(0)} \text{supp}(\alpha)} \cap \overline{\U_{\alpha \in \mcal N(\hat c)} \text{supp}(\alpha)} = \varnothing\]
    then by Urysohn's metricization lemma there exists a continuous separator $g: A \to [0, 1]$ such that $g$ is $0$ on $\overline{\U_{\alpha \in \mcal N(0)}} \text{supp}(\alpha)$ but $1$ on $\overline{\U_{\alpha \in \mcal N(\hat c)} \text{supp}(\alpha)}$. For fixed $\xi > 0$ and $\rho \in [0, 1 - \xi]$, define 
    \[ f(a, c') = f(a) = \rho + (1 - \xi - \rho) g(a). \]
    Note $f$ is continuous, independent of $c$, equals $\rho$ on every equilibrium where the short-run players' choose an action in $\mcal N(0)$, and equals $1 - \xi$ whenever short-run players choose $\mcal N(\hat c)$. The first property ensures that Assumption \ref{ass: dominance} is inherited by Condition (1) in the statement. From here, define 
    \[ \und v_{\hat c} = \min_{\alpha \in \mcal N(\hat c)} v(\alpha, \hat c), \text{ } \bar v = \sup_{a, c} v(a, c) \text{  and  } D = \sup_{a, c} \{ v(a, 0) - v(a, c) \}.  \]
    Note that we have 
    \[ V_* \geq \frac{(1 - \xi)\und v_{\hat c}}{\xi} \]
    since $\hat c$ is one action the long-run player could have committed to. Choose $\xi$ small enough that $\frac{(1 - \xi)\und v_{\hat c}}{\xi} > \max\{M, D + 1\}$. Note that $\bar V^0 \leq \frac{\rho \bar v}{1 - \rho}$; as $\rho$ vanishes so must $\bar V^0$. Thus, for sufficiently small $(\rho, \xi)$, we have first by construction and then by the definition of $D$ that 
    \[ V_* - \bar V^0 > D \geq v(\alpha, 0) - v(\alpha, c^*)\]
    which implies $c^*$-incentive compatibility since $f$ is independent of $c$. This gives the existence of a high-payoff equilibrium in which all long-run player types pool on $c^*$, guaranteeing $\bar U \geq V_* \geq M$. 

    For the low payoff equilibrium, set $\beta = 1 - \mu_0(\omega_R)$. Let 
    \[ F(\alpha) = \int_A f(a) d\alpha(a) \]
    be the average survival probability of an action profile $\alpha$. Since $\mcal N$ is upper hemi-continuous, one has that 
    \[ \lims_{\beta \to 0} \sup_{\alpha \in \mcal N(\beta c^*)} |F(\alpha) - \rho| \to 0. \]
    From here, consider the following strategy profile. 
    \begin{enumerate}
        \item In the first period, the rational long-run player plays $0$ and short-run players best reply with some $\alpha \in \mcal N(\beta c^*)$. 
        \item In all future periods, play $V_*$, regardless of the long-run player's actions. 
    \end{enumerate}
    That there exists a continuation profile yielding $V_*$ follows from the first part of the proposition along with the existence of a public randomization device. That the long-run player chooses to play $0$ in the first period follows from Assumption \ref{ass: dominance}. Finally, we have that the payoff is bounded by 
    \[ \sup_{\alpha \in \mcal N(\beta c^*)} F(\alpha)(\bar v + V_*) \to \rho(\bar v + V_*) \]
    as $\beta \to 0$. Fixing $\xi$ and taking $\rho \to 0$ then ensures that $\rho(\bar v + V_*) < \varepsilon$, which implies 
    \[ \limsup_{\mu_0(\omega_R) \to 1} \und U \leq \varepsilon \]
    as desired. This finishes the proof. 
\end{proof} 

\subsection*{PROOF OF LEMMA \ref{l: sequential learning}}
\label{sequential learning proof}
\begin{proof}
Let $(\sigma, \mu)$ be a sequential equilibrium and let $(\sigma^n, \mu^n)$ be the fully supported strategy which converges to it. Now fix any history $h^t$ where the long-run component is not $\{c^*\}_{s = 0}^{t - 1}$ noting it is supported by $\bb{P}^{\sigma^n}$ for all $n$. It must be that $\mu^n(h^t)(\omega_R) = 1$, since no committed type could have produced this history but there is positive probability the rational type did. This implies $\lims_{n \to \infty} \mu^n(h^t)(\omega_R) = \mu(h^t)(\omega_R) = 1$, finishing the proof. 
\end{proof}

\subsection*{PROOF OF LEMMA \ref{l: markov two actions}}
\label{markov two actions proof}
\begin{proof}
Let $(\sigma, \mu)$ be Markov perfect. By Proposition \ref{p: markov folk theorem}, any continuation strategy at histories $h^t$ where $\mu(h^t)(\omega_R) = 1$ must be a repeated static Nash profile, so in particular $\sigma_L(h^t) = 0$. 
Consider now a history $h^t$ where $\mu(h^t)(\omega_R) < 1$ (noting these are all on path), and suppose the lemma is false. Then there exists some on-path continuation history $h^t$ where at $h^t \succ h^{t - 1}$, $h^{t - 1}$ has long run component $\{c^*\}_{s = 0}^{t - 2}$ but $h_t = c \not\in \{0, c^*\}$. Thus, $\mu(h^t)(\omega_R) = 1$ and the long-run player's continuation value at $h^t$ must be the same as their continuation value at $\{h^{t - 1}, 0\}$. But then playing $0$ instead of $c$ for sure is a strictly profitable deviation by Assumption \ref{ass: dominance}, contradicting sequential rationality of $(\sigma, \mu)$. 
\end{proof}
 
\subsection*{PROOF OF LEMMA \ref{l: markov no mixing}}
\label{markov no mixing proof}
\begin{proof}
Two preliminary remarks first. First, define 
\[ \mcal N_\eta = \U_{\beta \in [1 - \eta, 1]} \mcal N(\beta c^*) \text{  and  } V_*(\eta) = \min_{\alpha \in \mcal N_\eta} \frac{f(\alpha, c^*) v(\alpha, c^*)}{1 - f(\alpha, c^*)}.\]
Then by compactness of the graph of $\mcal N$ and upper hemicontinuity, one has that $V_*(\eta) \to V_*$ as $\eta \to 0$. 
Second, under Assumption \ref{ass: c* IC}, there exists $\varepsilon > 0$ such that 
\[ f(\alpha, c^*)[v(\alpha, c^*) + V_*] \geq f(\alpha, 0)[v(\alpha, 0) + \bar V_0] + \varepsilon \text{  for all } \alpha \in \U_{\beta \in [0, 1]} \mcal N(\beta c^*)\]
where we use the strictness of $c^*$-incentive compatibility. 

Suppose not. Let $(\sigma, \mu)$ be an MPE and let $h^t$ be a history at which the long-run player mixes between $0$ and $c^*$. By the argument in Lemmas \ref{l: sequential learning} and \ref{l: markov two actions}, this can only possibly occur on a history $h^t$ where the long-run component is $\{c^*\}_{s = 0}^{t - 1}$ for some $t$; call such a history the time-$t$ clean history.
Define $\mcal T(h)$ to be the set 
\[ \{t : h^t \text{  is a time-$t$ clean history and } \sigma_L(h^t) \text{ mixes } \text{  and places probability at least } \eta \text{ on } 0. \}\]
Now fix some $\varepsilon$ from the second fact above, and choose $\eta$ from the first fact such that $|V_* - V_*(\eta)| < \varepsilon$.

First, $|\mcal T(h)|$ cannot be infinite for any $h^t$. Suppose not, so there is a sequence of times $\{t_n\}_{n = 1}^\infty \in \mcal T$ where the long-run player mixes with probability at least $\eta$ always. 
Note that when the long-run player mixes with probability bounded away from $0$, both (1) $h^{t_n} \in \bb{P}^\sigma$ where $h^{t_n}$ is clean, and (2) the long-run player's action is statistically identifiable compared to the committed type. This implies that players eventually learn the long-run player's type; (see \cite{mailath2006repeated} Chapter 15) that is, $\lims_{n \to \infty} \mu(h^{t_n})(\omega_{c^*}) = 1$ almost surely. Because short-run player actions are upper hemi-continuous in $\mu(h^{t_n})$, there must exist $N$ sufficiently large that for all $n \geq N$, $\mcal V(h^{t_n} | \sigma) \geq \mcal V(c^*, h^{t_n}) \geq V_*(\eta)$, where $\mcal V(c^*, h^t)$ is the long-run player's continuation history from simply always taking $c^*$ in the future and $\mcal V(h^t | \sigma)$ is their continuation value from $\sigma$. The second inequality follows since the unconditional strategy at every clean history is of the form $\beta c^*$ and we can choose the mixing time such that $\beta_t$ such that $\beta_t \geq 1 - \eta$, so every stage-game pair lies in $\mcal N_\eta$. 
Yet mixing implies the long-run player must be indifferent between taking $0$ (and obtaining at most $\overline V_0$ hereafter, since they have revealed their rationality) and taking $c^*$ (and obtaining some other future continuation value). In particular, for $N$ sufficiently large and $a = \{\sigma_i(h^{t_N})\}$, we have  
\[ f(a, 0)[v(a, 0) +  \mcal V(\{h^{t_n}, 0\} | \sigma)] = f(a, c^*)[v(a, c^*) + \mcal V(\{h^{t_n}, c^*\} | \sigma)]. \]
Yet we also know from Proposition \ref{p: markov folk theorem} that for all $\varepsilon > 0$, there exists $N$ sufficiently large such that 
\[  f(a, 0)[v(a, 0) + \mcal V(\{h^{t_n}, 0\} | \sigma)] \leq  f(a, 0)[v(a, 0) + \overline V_0] + \varepsilon\]
and (combined with from our observation above) that 
\vspace{-0.4em}
\[ f(a, c^*)[v(a, c^*) +  \mcal V(\{h^{t_n}, c^*\} | \sigma)] 
\geq f(a, c^*)[v(a, c^*) + V_*(\eta)]. \]
\vspace{-0.4em}
Yet the assumption the long-run player satisfies $c^*$ incentive compatibility implies 
\[ f(a, 0)[v(a, 0) +  \overline V_0] < f(a, c^*)[v(a, c^*) +  V_*(\eta)] \]
for $a$ since $a \in \mcal N(\beta c)$. Taking everything together and using $c^*$-incentive compatibility implies for all $N$ sufficiently large and $\varepsilon$ sufficiently small 
\[  f(a, 0)[v(a, 0) +  \mcal V(\{h^t, 0\} | \sigma)] < f(a, c^*)[v(a, c^*) + \mcal V(\{h^{t_n}, c^*\} | \sigma)]\]
contradicting the indifference condition which is necessary to mix. 

Thus, $|\mcal T(h)|$ is finite. Let $T > 0$ be the last time in $\mcal T(h)$ for some history. This implies after $T$, the long-run player attains at least $V_*(\eta)$ when they play $c^*$ forevermore, by our choice of $\eta$. Thus, 
\begin{align*}
    f(a, c^*)[v(a, c^*) + V_*(\eta)] \geq f(a, c^*)[v(a, c^*) + V_*] - (1 - \xi)\varepsilon 
    \\ \geq f(a, 0)[v(a, 0) + \bar V_0] + \xi \varepsilon > f(a, 0)[v(a, 0) + \bar V_0]
\end{align*}
using the fact $f \leq 1 - \xi$. This contradicts the indifference condition required for mixing and hence $T(h) = \varnothing$. Thus at every clean history the rational type puts weight below $\eta$ on $0$, so 
\[ \beta_t = \mu_t(\omega_{c^*}) + \mu_t(\omega_R) \sigma_t(c^*) \geq 1 - \eta\]
at all clean histories, so every stage game pair lies in $\mcal N_\eta$, and our computations above apply at any mixing history. Hence there can be no mixing anywhere. 
\end{proof}

\subsection*{PROOF OF THEOREM \ref{t: commitment theorem} (Converse)}
\label{converse commitment theorem proof}
\begin{proof}
Suppose the condition fails. There are three cases. First, suppose the inequality fails strictly at $\beta = 0$, but holds at $\beta = 1$. Then in particular 
\[ f(a, 0)[v(a, 0) +  \overline V_0] > f(a, c^*)[v(a, c^*) +  V_*] \]
for $a \in \mcal N(0)$. 
This IC constraint implies that the rational long-run player always prefers to separate and play $0$ (starting from any time-$0$ belief) if they are guaranteed $\overline V_0$ afterwards. Thus, since $\mcal N(0)$ is single-valued and upper hemi-continuous, there exists $\bar \mu_0 < 1$ such that the following is a Markov equilibrium for all $\mu_0(\omega_R) > \bar \mu_0$:
\begin{enumerate}
    \item The rational long-run player plays $0$ in every period on the path of play where $\mu_0(\omega_R) > 0$.
    \item Short-run players play $a_* = a^*(c^*)$ at all $h^t$ where $\mu(h^t)(\omega_R) = 0$, with $a_*$ chosen to deliver the long-run player expected continuation payoff $V_*$. Long-run players play $c^*$.
    \item If $\mu(h^t)(\omega_R) = 1$, play $a_0^* \in \mcal N(0)$ chosen to deliver the long-run player an expected continuation payoff $\overline V_0$. 
    \item At histories where $\mu(h^t)(\omega_R) = \mu_0(\omega_R)$, play any $a \in \mcal N(\mu_0(\omega_{c^*})c^*)$. 
\end{enumerate}
Second, suppose instead that the inequality fails for some $\beta \in (0, 1)$ but that it holds strictly for $\beta = 1$. This implies first that it is possible to sustain the Markov equilibrium where the long-run player plays $c^*$ at all histories.
Moreover, there must exist some $\overline \beta \in (0, 1)$ with $a \in \mcal N(\beta c^*)$ and
\[ f(a, 0)[v(a, 0) +  \overline V_0] = f(a, c^*)[v(a, c^*) +  V_*]. \]
To see this, let 
\[ \phi(\beta) = \min_{\alpha \in \mcal N(\beta c^*)} \underbrace{\left[ f(\alpha, c^*)[v(\alpha, c^*) + V_*] - f(\alpha, 0)[v(\alpha, 0) + \bar V_0] \right]}_{g(\alpha)} .\]
Let $\hat \beta = \sup\{\beta : \phi(\beta) < 0\}$ and note that by upper hemi-continuity of $\mcal N$ (taking $\beta_n \to \hat \beta$ as necessary, and $\alpha_n \in \mcal N(\beta_n)$), we can find some $\alpha^+, \alpha^-$ such that $g(\alpha^+) \leq 0$ but $g(\alpha^-) \geq 0$. Since $g$ is continuous and $\mcal N(\hat \beta c^*)$ is connected, we get that there is some $a \in \mcal N(\hat \beta, c^*)$ such that $g(a) = 0$, as desired. 
This implies that there exists a Markov equilibrium of the following form:
\begin{enumerate}
    \item If $\mu(h^t)(\omega_R) \not\in \{\mu_0(\omega_R), 1\}$, then have $\sigma_L(h^t) = c^*$ and $\{\sigma_i(h^t)\} \in \mcal N(c^*)$ be chosen to give the long-run player $V_*$. 
    \item If $\mu(h^t)(\omega_R) = 1$, play static Nash $\sigma(h^t)$ giving the long-run player $\overline V_0$.
    \item If $\mu(h^t)(\omega_R) = \mu_0(\omega_R)$, have $a = \{\sigma_i\} \in \mcal N(\beta c^*)$ chosen so that 
   \[ f(a, 0)[v(a, 0) +  \overline V_0] = f(a, c^*)[v(a, c^*) +  V_*] \]
   and $\sigma_L(h^t) \in (0, 1)$ chosen so that 
   \[ \mu_0(\omega_R) \sigma(h^t)(c^*) + (1 - \mu_0(\omega_R)) = \beta \]
   Noting this is well-defined so long as $\mu_0(\omega_R) > 1 - \beta$. 
\end{enumerate}
 It is clear that at all histories, the short-run player is best replying, as is the long-run player at all histories outside of $\varnothing$. Moreover, at the first history $h^t = \varnothing$, by choice of $\beta$, the long-run player is indifferent between $c^*$ and $0$ and thus is willing to randomize given $a$. Hence this is a Markov perfect equilibrium. This gives the long-run player an equilibrium payoff of 
\begin{align*}
    f(a, 0)[v(a, 0) +  \overline V_0] < f(a_*, 0)[v(a_*, 0) +  \overline V_0]
    \\ \leq f(a_*, c^*)[v(a_*, c^*) +  V_*]
\end{align*}
where the first inequality follows from monotonicity of the game and the second follows from the fact the incentive compatibility condition holds at $\beta = 1$ where $a_* = a^*(c^*)$ is the largest element of $\mcal N(c^*)$, since again by the monotonicity assumption this is the choice which minimizes the stage game payoff. Note in particular $a \succ a_*$because the game is monotone. 
This implies the result. 

Finally, suppose the inequality fails when $\beta = 1$. Then there cannot be an equilibrium where the long-run player imitates $c^*$ for sure. The only other pure Markov equilibrium gives the long-run player a payoff at most $f(a, 0)[v(a, 0) + \bar V_0]$ for $a \in \mcal N(\mu_0(\omega_{c^*})c^*)$, which delivers the result. If instead the long-run player mixes in a Markov equilibrium at some on-path history, they secure (for some $a \in \mcal N(\beta c^*)$ and some $\beta \in (0, 1)$) a payoff of at most 
\[ f(a, c^*)[v(a, c^*) +  V_*] < f(a^*, c^*)[v(a^*, c^*) +  V_*] = V_* \]
by monotonicity and hence must secure a payoff smaller than $V_*$. 
\end{proof}

\subsection*{PROOF OF PROPOSITION \ref{p: chain store game}}
\label{p: chain store game proof}
\begin{proof}
Write $F$ for Fight and $A$ for Accommodate. If $\beta$ is the probability assigned to $F$, a computation gives that 
\[ \mcal N(\beta F) = \begin{cases} \{In\} & \text{  if  }  \beta < \frac12
 \\ \Delta(\{In, Out\}) & \text{  if  } \beta = \frac12 \\ 
 \{Out\} & \text{  if  } \beta > \frac12 \end{cases}. \] 
Since $\delta \in (0, 1)$, one can verify that Assumption \ref{ass: dominance} holds. From here, note that the commitment payoff is exactly $K = \frac{(1 - \eta)(2 - \delta)}{\eta}$; meanwhile, if the incumbent chooses $A$ and the entrant plays In, the payoff is $V^0(\varepsilon) = \frac{2\varepsilon}{1 - \varepsilon}$. Thus $F$ is the Stackelberg action if and only if 
\[ K > V^0(\varepsilon) \iff \varepsilon < \varepsilon_s = \frac{K}{K + 2}. \]
Since moreover $\varepsilon_s < 1 - \eta$ since $\delta > 0$, we have that for any $\varepsilon < \varepsilon_s$, the game is monotone with $In \succ Out$. 

Now the equilibrium constructions. Fix $\varepsilon < \varepsilon_s$; $\mcal N(A) = \{In\}$, and so $\bar V^0 = V^0(\varepsilon)$. When the entrant plays In, the incentive constraint is 
\[ \varepsilon K > \varepsilon(2 + V^0(\varepsilon) \iff K > \frac{2}{1 - \varepsilon} \iff \varepsilon < \bar \varepsilon = 1 - \frac{2}{K}. \]
Note that $K > 2$ since $\eta \in (0, \frac13]$ and $\delta \in (0, 1)$, so $\bar \varepsilon \in (0, 1)$ and $\varepsilon_s > \bar \varepsilon$. 
When the entrant plays Out, the incentive constraint gives that 
\[ (1 - \eta)((2 - \delta) + K) > \left(1 - \frac\eta2\right)(2 + V^0(\varepsilon)).\] 
Note that $2 + V^0(\varepsilon)$ is increasing in $\varepsilon$; when $\varepsilon = \bar \varepsilon$, we have that 
\[ 2 + V^0(\bar \varepsilon) = \frac{2}{1 - \bar \varepsilon} = K\]
and so the Out constraint simplifies to requiring that $K > (1 - \frac{\eta}{2})K$, which is satisfied strictly. Thus the Out IC is satisfied for all $\varepsilon \leq \bar \varepsilon$. 

Thus strict $F$-incentive compatibility holds for all $\varepsilon < \bar \varepsilon$, and Theorem \ref{t: commitment theorem} implies each Markov equilibrium has the rational incumbent play $F$ on path, giving $U^M = V_* = K$. 

If instead $\varepsilon \in (\bar \varepsilon, \varepsilon_s)$, the In constraint fails, and thus by monotonicity the converse to Theorem \ref{t: commitment theorem} implies the desired statement. 

Finally, fix $\varepsilon < \bar \varepsilon$. Consider the following strategy profile, which is a sequential equilibrium for all $\mu_0(\omega_R) > \frac12$. 
\begin{enumerate}
    \item In period $0$, the rational incumbent plays $A$, and the entrant chooses In. 
    \item Following either incumbent action, pool on $F$ forever and give the long-run player a payoff of $K$. 
\end{enumerate}
Because in the first period the incumbent is unlikely to survive, the equilibrium payoff is at most $\varepsilon(2 + K)$. This proves the last argument. 
\end{proof}

\subsection*{PROOF OF PROPOSITION \ref{p: stage game properties}}
\label{p: stage game properties proof}
\begin{proof}
Fix a value $c$, and note the induced stage game is a finite monotone (decreasing) supermodular game with dominance regions $[-\infty, 0)$ and $(1, \infty]$. This implies by the proof of Theorem 1 of \cite{MorrisPauznerFrankel03} that for all $\varsigma$ sufficiently small, $s(x, c) = \mathbf{1}\left\{x < x^*(c)\right\}$ is the unique equilibrium strategy profile. 

Next, let $U_i(c, x, x^*)$ be the interim utility differential between taking $a = 1$ and $a = 0$ for an agent observing signal $x$ when players use a cutoff strategy $x^*$ and the cost of failed revolution is $c$. In equilibrium, it must be that $U_i(c, x^*(c), x^*(c)) = 0$.
Regime change only occurs if $F_x(x^*(c) | \theta) > \theta$, so $\theta^*(c)$ is implicitly defined by the equation $F_x(x^*(c) | \theta) - \theta = 0$. Differentiating yields
\[ \frac{d}{d\theta} (F_x(x^*(c) | \theta) - \theta) = \frac{\partial}{\partial 2} F_x(x^*(c) | \theta) - 1 < 0\]
as $\frac{\partial}{\partial 2} F_x(x | \theta) < 0$ for all $x$; this then implies $\theta^*(c)$ is globally $\mcal C^1$. Applying the implicit function theorem again implies its derivative is given by  
\[ \frac{\partial \theta^*(c)}{\partial c} = -\left(f_x(x^*(c) | \theta) \frac{\partial x^*(c)}{\partial c}\right)\left( \frac{\partial}{\partial 2} F_x(x^*(c) | \theta) - 1 \right)^{-1}, \]
which is negative so long as $x^*(c)$ is decreasing. For this, note that 
\[ U_i(c', x, y) - U_i(c, x, y) = (c' - c)[F_\theta(\theta^*(y) | x) - 1] < 0\]
pointwise. Setting $\Phi(c, x) = U_i9c, x, x)$, we have that $\Phi(c', x^*(c)) < \Phi(c, x^*(c))$ Since $\Phi(c', \cdot)$ is positive below the lower dominance region and has a unique zero, that zero lies strictly below $x^*(c)$. Hence $x^*(c') < x^*(c)$. This then implies that higher costs to failed attacking shrink the region of states where such attacks are successful. 
\end{proof}

\subsection*{PROOF OF THEOREM \ref{t: global games theorem}}
\label{t: global games theorem proof}
\begin{proof}
Fix some distribution $P_\theta$. Note  
\[ \lims_{\varsigma \to 0} F_x\left(\frac{1}{1 + c^*} \bigg | x \right) \to \begin{cases} 1 \text{  if  }  x < \frac{1}{1 + c^*} \\ 0 \text{  if  } x > \frac{1}{1 + c^*} \end{cases} \]
because $F_x(\cdot | \theta)$ converges to $\delta_\theta$ as $\varsigma \to 0$ (people become extremely sure about the state). Since $\theta^*(c)$ is continuous and zeros $F_x\left(\frac{1}{1 + c^*} | \theta\right) - \theta$, this implies $\theta^*(c) = \frac{1}{1 + c}$. 

The first statement. Suppose $\frac{1}{1 + c^*} < \mu$ and let $\eta = \mu - \frac{1}{1 + c^*}$, noting this gap is independent of $\tau$. We want to prove
\[ V^* - \und V_0  > \kappa(c^*) \]
in the right order of limits. Set $\overline P_\theta(x) = 1 - P_\theta(\theta^*(x))$ to be the probability of survival given a commitment posture. Some algebra implies that incentive compatibility is equivalent to the requirement 
    \[ \overline P_\theta(c^*) - \overline P_\theta(0) \geq \kappa(c^*)[1 - \overline P_\theta(0)]. \]
    Note from here that  
    \[ \overline P_\theta(c^*) = 1 - P_\theta(\theta^*(c^*)) > 1 - P_\theta(\mu - \eta) \geq 1 - \frac{\var(\theta)}{\eta^2} > 1 - \frac{\overline \tau}{\eta^2}. \]
The middle inequality uses Chebychev's inequality. Thus $\overline P_\theta(c^*) \to 1$ as $\overline \tau \to 0$; similarly $\overline P_\theta(0) \to 0$ as $\overline \tau \to 0$; together this implies the incentive constraint holds. Theorem \ref{t: commitment theorem} implies the regime can sustain $c^*$ and some best response to $c^*$ in the unique Markov equilibrium, as desired. Finally because $\overline P_\theta(c^*) = f(c^*)$, the exit probability vanishes as $\overline \tau \to 0$. This implies both parts of the first statement, noting $\overline \varsigma$ can depend on the choice of $\tau$. 

   Now consider the case where $\frac{1}{1 + c^*} > \mu$, now setting $\eta = \frac{1}{1 + c^*} - \mu > 0$. Note a necessary condition for incentive constraints to hold is for 
   \[ \frac{\overline P_\theta(c^*) - \overline P_\theta(0)}{1 - \overline P_\theta(0)} > \kappa(c^*)\]
  As $\bar \tau$ vanishes, one has that 
   \[ 1 - P_\theta(\theta^*(\sigma^*(h^t))) \leq 1 - P_\theta(\theta^*(c^*)) < 1 - P_\theta(\mu + \eta) \leq \frac{\var(\theta)}{\eta^2} < \frac{\overline \tau}{\eta^2} \]
   again using Chebychev's inequality. Since $\theta^*(c) < \theta^*(0)$, we conclude that 
   \[ \frac{\overline P_\theta(c^*) - \overline P_\theta(0)}{1 - \overline P_\theta(0)} \leq \frac{\frac{\bar \tau}{\eta^2} - \overline P_\theta(0)}{1 - \overline P_\theta(0)}.  \] 
   Hence this term is bounded from above by $\kappa(c^*)$ for all sufficiently small $\tau$. 
   From here, note that any per-period survival probability is bounded by $\frac{\tau}{\eta^2}$, and so any payoff stream is bounded by $\frac{p}{1 - p}$ with $p \leq \frac{\tau}{\eta^2}$. Taking $\tau \to 0$ delivers the bound. 
\end{proof}

\newpage 
\section*{SUPPLEMENTAL APPENDIX}

\appendix

\makeatletter\def\@currentlabel{Appendix B.}\makeatother
\label{Appendix B}

\setcounter{lemma}{0}
\renewcommand{\thelemma}{B.\arabic{lemma}}

\setcounter{proposition}{0}
\renewcommand{\theproposition}{B.\arabic{proposition}}

\setcounter{definition}{0}
\renewcommand{\thedefinition}{B.\arabic{definition}}

\setcounter{corollary}{0}
\renewcommand{\thecorollary}{B.\arabic{corollary}}

This appendix considers the discounted repeated game at some discount rate $\delta$. Section B.1 discusses some problems with normalized repeated games, interpreted as the average payoffs across periods. Section B.2 derives a novel comparative statics result on the set of commitment postures as the discount rate changes, and interprets it in the context of applications. Section B.3 sets the main result in the language of the discounted game. 

\subsection*{B.1. Non-normalized payoffs}

Intuitively, without exit, the stream of normalized discounted payoffs converges to the average payoff as $\delta \to 1$. However, when the game has endogenous exit, so long as exit occurs with high enough probability, the average stream of payoffs will be zero for ``most'' periods (i.e. if the expected exit time is finite), so normalized payoffs will vanish as players get extremely patient. The lemma formalizes this observation. 

\begin{lemma}
\label{l: discounted payoffs}
    Let $\{f_t\} \subset [0, 1]$ be a sequence of exit probabilities. If $\liminf_{t \to \infty} f_t < 1$, then 
    \[ \lims_{\delta \to 1} (1 - \delta)\sum_{t = 0}^\infty \delta^t v_t \prod_{s= 0}^t f_s = 0. \]
    for any feasible payoffs $v_t$. 
\end{lemma}
\begin{proof}
 Let $\{\overline f_{\tilde t}\} \subset \{f_t\}$ be a subsequence such that $\lims_{\tilde t \to \infty} \overline f_{\tilde t} = \overline f < 1 - \eta$, for some $\eta > 0$ fixed; moreover, by truncating finitely many terms at most, suppose $\overline f_{\tilde t} < 1 - \eta$ for all $s \in \bb{N}$. Now for any $T \in \bb{N}$, define the number of times that $f_t < 1 - \eta$ to be $s(T) = \#\{s : f_s < 1 - \eta\}$. By assumption of $\eta$, we know $\lims_{T \to \infty} s(T) = \infty$. 
 We now have that $\prod_{s = 0}^\infty f_t < (1 - \eta)^{s(T)}$. Letting $\overline v$ be chosen so that $|v_t| \leq \overline v$ for all $t$, we have 
 \begin{align*}
     0 & \leq \lims_{\delta \to 1} \left| (1 - \delta)\sum_{t = 0}^\infty \delta^t v_t \prod_{s= 0}^\infty f_s \right|  \leq \lims_{\delta \to 1} \left|(1 - \delta)\sum_{t = 0}^\infty  \delta^t \overline v (1 - \eta)^{s(t)} \right|
     \\ & \leq \lims_{\delta \to 1}  \left|(1 - \delta) \sum_{t = 0}^T  \delta^t \overline v (1 - \eta)^{s(t)}\right| + \left|(1 - \delta)(1 - \eta)^{s(T)}\sum_{t = T + 1}^\infty \delta^t \overline v \right|
     \\ & \leq (1 - \eta)^{s(T)} \lims_{\delta \to 1} (1 - \delta) \sum_{t = T + 1}^\infty \delta^t \overline v  \leq (1 - \eta)^{s(T)} \overline v. 
 \end{align*}
 Since this holds for all $T \in \bb{N}$, taking $T \to \infty$ (and hence $s(T) \to \infty$) gives the argument. 
\end{proof}

Lemma \ref{l: discounted payoffs} implies that it is not fruitful to compare discounted average payoffs in the patient limit, as they will often be zero even when payoffs are meaningfully different. 

There are two potential ways to ameliorate this problem. The first is to normalize by the \emph{effective discount rate}, $f(a, c)\delta$, in each period. However, such a normalization is infeasible and difficult to interpret: $f(a, c)$ is an endogenous object, and thus one should naturally expect that it vary the long-run players' payoffs. Moreover, except in special cases, the stream of exit probabilities $\{f(a_t, c_t)\}_{t = 0}^\infty$ need not be time-invariant, making it hard to understand what the ``average effective discount rate'' should be. 
The second way is to simply work directly with the undiscounted payoffs $\mcal V(\{a_t, c_t\})$; I take this approach in this paper. 

This choice, however, leads itself to a novel technical challenge: without normalizing payoffs by the discount rate, deviations today may remain first order (in particular, they are not order $(1 - \delta)$ compared to an order $\delta$ variation in the continuation payoff, and are instead order $1$ versus order $\delta$ variation in the continuation payoff). This makes it much harder to discipline the long-run player from deviating to play non-Stackelberg actions, since deviations from potential equilibrium profiles will always remain tempting. 
Theorem \ref{t: commitment theorem} can thus be interpreted as providing a simple economic condition under which the long-run player can successfully establish a (Markov) reputation, despite this hurdle.

\subsection*{B.2. Commitment Postures}
How does the discount rate affect the long-run player's Stackelberg payoff? Define 
\[ c^*(\delta) = \argmax_{c \in C} \min_{a \in \mcal N(c)} \sum_{t = 0}^\infty \delta^t f(a, c)^{t + 1} v(a, c) \]
to be the set of (lower) Stackelberg actions for the long-run player when they discount the future at rate $\delta$. Let $V^*(c, \delta)$ be the value from playing the lower Stackelberg action. 
Note the long-run player trades off between a higher {survival probability} at the (potential) cost of a lower {stage game payoff} in each period. My next result characterizes exactly how the long-run player resolves this tradeoff as they get more patient: regardless of the speed at which $v(a, c)$ might decline in $c$, they will always favor more aggressive actions that increase their survival probability as they get more patient. To do so, I make one more assumption on the survival probability.

\begin{assumption}
\label{ass: survival probability}
    The correspondence $\und a(c) = \argmin_{a \in \mcal N(c)} \sum_{t = 0}^\infty \delta^t f(a, c)^{t + 1} v(a, c)$ is single-valued, and $f^*(c) = f(\und a(c), c)$ is strictly increasing in $c$. 
\end{assumption}

\begin{proposition}[Patience Effect]
\label{p: patience effect}
    Suppose Assumption \ref{ass: survival probability} is satisfied. $c^*$ is increasing (in the strong set order) in $\delta$. 
\end{proposition}
\begin{proof}
  Monotonicity now relies on the following lemma.\footnote{Such a lemma appears elementary, but to the best of my knowledge there is not a prior reference. Note the lemma establishes a general comparative statics result for discounted streams of payoffs that cross zero once from below.}
   \begin{lemma}
   \label{l: twitter question}
    Let $\{b_t\}_{t \in \bb{N}} \subset \bb{R}$ be a bounded sequence of real numbers, not identically zero, such that $b_t < 0$ if and only if $t < T$ for some $T \in \bb{N}$. Then for any $\delta_1 < \delta_2$,  
    \[ \sum_{t = 0}^\infty \delta_1^t b_t \geq 0 \implies \sum_{t = 0}^\infty \delta_2^t b_t > 0. \]
\end{lemma}
    \begin{proof}
    Note 
    \[  \sum_{t = 0}^\infty \delta_1^t b_t =  \delta^T \sum_{t = 0}^\infty \delta_1^{t - T} b_t \geq 0 \iff \sum_{t = 0}^\infty \delta_1^{t - T} b_t \geq 0  \]
As a function of $\delta$, this series is absolutely convergent for all $\delta \in (0, 1)$ and the partial sums converge uniformly on this interval so we can differentiate under the summation: thus, 
    \[ \frac{d}{d \delta} \sum_{t = 0}^\infty \delta_1^{t - T} b_t = \sum_{t = 0}^\infty (t - T) \delta_1^{t - T - 1} b_t < \infty \]
    where we use the fact $b_t$ is bounded to guarantee the sum is well-defined. Note each term in this summation is nonnegative. 
    If $t < T$, then $t - T < 0$ and $b_t < 0$, so $(t - T)\delta^{t - T - 1} b_t > 0$, while if $t > T$, then $(t - T) \delta^{t - T - 1} \geq 0$ and $b_t \geq 0$ as well. Moreover, since $\{b_t\}$ is not identically zero, at least one term in the sum above must be strictly positive, and hence the derivative is strictly positive. We thus have that 
    \[ \sum_{t = 0}^\infty \delta_2^{t - T} b_t > 0 \implies \delta_2^T \sum_{t = 0}^\infty \delta_2^{t - T} b_t = \sum_{t = 0}^\infty \delta_2^t b_t > 0 \]
    where we use the fact $\delta_2 > \delta_1$, and finally multiply by $\delta_2^T > 0$. 
\end{proof}
Fix any $\delta$ and $c' > c$; by Assumption \ref{ass: survival probability}, it must be that for all $(c', c)$, $f^*(c'), f^*(c) > 0$. 
Consider the difference 
\[ \sum_{t = 0}^\infty  \delta^t [ v(a^*(c'), c') f^*(c')^{t + 1} - v(a^*(c), c)f^*(c)^{t + 1}] \]
where I abuse notation to and have  $a^*(\cdot)$ be any selection from the minimizing correspondence. Note that if $b_t$ is defined as
\[ b_t = v(a^*(c'), c') f^*(c')^{t + 1} - v(a^*(c), c)f^*(c)^{t + 1}  \]
then 
\[ b_t \geq 0 \iff \frac{v(a^*(c'), c')}{v(a^*(c), c)} \geq \left(\frac{f^*(c)}{f^*(c')}\right)^{t + 1}.\]
The left hand side is strictly positive, while the right hand side is positive, in $(0, 1)$ (by Assumption \ref{ass: survival probability}), and strictly decreasing in $t$. Thus, there exists a unique $T$ (possibly $0$) at which the inequality holds for all $t \geq T$. Lemma \ref{l: twitter question} then implies, for any $c' > c$, $\sum_{t = 0}^\infty \delta^t b_t$ is increasing in $\delta$, and hence the objective function 
$\sum_{t = 0}^\infty  v(a^*(c), c)$ 
is single-crossing in $(\delta, c)$. The result then follows by the \cite{MilgromShannon94} comparative statics theorem. 
\end{proof}

Proposition \ref{p: patience effect} applies regardless of the speed or curvature of $v$ (in particular, it holds even if $v$ declines very fast relative to $c$), though this comes at the cost of a weak comparative static (every selection from $c^*$ is nondecreasing in $c$). Moreover, it has economic implications in the context of the applications---a regime seeking to fend off a coordinated attack will invest more in survival when they are more patient; longer-lived firms care more about ensures entrants are deterred; patient motorcyclists have a greater incentive to take ``separating'' actions that prove they are not commitment types, in order to deter exit.

\subsection*{B.3. The Main Result}
That $c^*$ varies with the discount rate produces a new order of limits challenge in trying to prove some version of a commitment payoff theorem in the discounted repeated game. In particular, since the long-run player's Stackelberg {action} depends on the discount rate, it is not clear that there is a finite set $\Omega$ such that $\Omega \cap c^*$ is nonempty for all $\delta$ close to $1$. Consequently, for any fixed set of commitment types, it may be infeasible to approximate a patient committed sender. 

To obviate this problem, I index and simplify the set of commitment postures, $\Omega = \{\omega_R, \omega_{c^*}\}$ (where I abuse notation and use $c^*$ to refer to a {single selection} from the set of commitment types). Moreover, as before, I set $\und U$ and $\overline U$ as the infimum and supremum of the long-run player's best and worst sequential equilibrium payoff, though now I implicitly fix the set of commitment types $\Omega = \Omega$, though I still suppress dependence on the specific sequence of prior beliefs $\mu_0 \in \Delta(\Omega)$ (which itself my also change in $\delta$). Given this notation, the limit infimum (supremum)
$\liminf_{\delta \to 1} \und U$ (resp. $\limsup_{\delta \to 1}$)
should be interpreted as the limit infimum (supremum) of the sequence of lower equilibrium payoffs in the repeated game with discount rate $\delta$, commitment types $\Omega$, and some fixed sequence of prior beliefs $\mu_0 \in \Delta(\Omega)^o$. 
This interpretation implies that statements of the form $\limsup_{\delta \to 1} |V_* - \und U| = 0$ can be read as ``The long-run player attains almost their Stackelberg payoff so long as there is positive probability players believe they are committed to their Stackelberg action \textit{at their level of patience}''. 

Under this interpretation, the following definition and result are immediate applications of the undiscounted game, by continuity of the payoff parameters in the discount rate. 

\begin{definition}
    Say the long-run player \emph{asymptotically satisfies (weak) $c^*$ incentive compatibility} if there exists $\overline \delta < 1$ such that for all $\delta > \overline \delta$ and each $\alpha \in \mcal N(\beta c^*(\delta))$ for $\beta \in [0, 1]$, 
    \[ f(\alpha, c^*(\delta))[v(\alpha, c^*(\delta)) + \delta V_*(\delta)] > (\geq) f(\alpha, 0)[v(\alpha, 0) + \delta \bar V_0(\delta)]\] 
\end{definition}

\begin{theorem}
   Let $\Omega(\delta) = \{\omega_R, \omega_{c^*(\delta)}\}$. If the long-run player  asymptotically satisfies $c^*$ incentive compatibility, then for all $\mu_0$ with $\mu_0(\omega_R) < 1$, 
   \[ \limsup_{\delta \to 1} |\und U^M(\delta) - V_*(\delta)| = 0. \]
   Conversely, if $(u, v, f)$ is monotone and $\mcal N(0)$ is single-valued, and the long-run player does not asymptotically satisfy weak $c^*$ incentive compatibility, then
   \[ \limsup_{(\delta, \mu_0(\omega_R)) \to (1, 1)} |\und U^M(\delta) - V_*(\delta)| > 0. \]
\end{theorem}

\end{document}